\documentclass[12pt,draftcls,onecolumn]{IEEEtran}

\IEEEoverridecommandlockouts

\usepackage{epsfig}
\usepackage{cite}
\usepackage{graphicx}
\graphicspath{{Figures/}}
\usepackage{url}
\usepackage{stfloats}
\usepackage{amsmath}
\usepackage{amsfonts}   
\usepackage{amsopn}     
\usepackage{psfrag}
\usepackage[tight]{subfigure}
\usepackage{mdwtab}     
\usepackage{enumerate}
\usepackage{color}

\newtheorem{theorem}{\it Theorem}
\newtheorem{proposition}{\it Proposition}
\newtheorem{rem}{\it Remark}

\newtheorem{lemma}{\it Lemma}
\newtheorem{example}{\it Example}

\begin{document}

\title{Node Classification in Networks of Stochastic Evidence Accumulators}

\author{Ioannis~Poulakakis~\IEEEmembership{Member; IEEE}, Luca Scardovi~\IEEEmembership{Member; IEEE}, \\ and Naomi Ehrich Leonard~\IEEEmembership{Fellow; IEEE}
%
%
\thanks{Manuscript submitted on October 16, 2012. Preliminary parts of this paper appear in \cite{PoScLe10,PoScLe12}.}
\thanks{I. Poulakakis is with the Department of Mechanical Engineering, University of Delaware, Newark, DE 19711, USA;  e-mail: {\tt\small poulakas@udel.edu.}}
\thanks{L. Scardovi is with the Department of Electrical and Computer Engineering, University of Toronto, Toronto, ON, M5S 3G4, Canada; e-mail: {\tt\small scardovi@scg.utoronto.ca.}}
\thanks{N. E. Leonard is with the Department of Mechanical and Aerospace Engineering, Princeton University, Princeton, NJ 08544, USA; e-mail: {\tt\small naomi@princeton.edu.}}
}


\markboth{I. POULAKAKIS, L. SCARDOVI, N. E. LEONARD}{NODE CLASSIFICATION IN NETWORKS OF STOCHASTIC EVIDENCE ACCUMULATORS}

\maketitle

\begin{abstract}
This paper considers a network of stochastic evidence accumulators, each represented by a drift-diffusion model accruing evidence towards a decision in continuous time by observing a noisy signal and by exchanging information with other units according to a fixed communication graph. We bring into focus the relationship between the location of each unit in the communication graph and its certainty as measured by the inverse of the variance of its state. We show that node classification according to degree distributions or geodesic distances cannot faithfully capture node ranking in terms of certainty. Instead, \emph{all} possible paths connecting each unit with the rest in the network must be incorporated.  We make this precise by proving that node classification according to information centrality provides a rank ordering with respect to node certainty, thereby affording a direct interpretation of the certainty level of each unit in terms of the structural properties of the underlying communication graph. 

\end{abstract}

%
\IEEEpeerreviewmaketitle

\newpage

\section{Introduction}
\label{sec:TAC10:intro}

\IEEEPARstart{I}{n} networks of sensors accumulating evidence by observing noisy processes, information sharing among the individual sensing units can significantly affect their certainty about the processes observed. Depending on the communication architecture, units that are more certain than others emerge, and these more certain units may prove to be more reliable decision makers in collective decision-making tasks \cite{PoScLe10}, or more influential components in consensus-seeking networks in the presence of noise \cite{YoScLe10, Med11}.

The identification of the most certain units in a network of interconnected systems is central to shaping collective behavior. For instance, teams of autonomous vehicles used as mobile sensor networks in the ocean \cite{Fiorelli06}, on land \cite{Gustafson05}, in the air and in space \cite{Mesbahi01, Richards2002}, must perform exploration, surveillance, monitoring, search and rescue, and manipulation tasks by responding quickly and accurately to noisy measurements of uncertain environmental processes \cite{Cao1997}. If it is understood which individuals are the most certain about their environment due to their location in the network, protocols could be adapted so that these individuals dominate the group's behavior, e.g. by suitably weighting the information supplied \cite{Couzin11}, or the decision made \cite{Dandach12IEEE}, by each unit.

The contribution of the present paper is to characterize the impact of the communication architecture on the quality of the information content of each unit in a network of stochastic evidence accumulators. In decision making, evidence accumulation often assumes that relevant information is collected sequentially, through a series of independent scalar observations. This assumption forms the basis for a large class of decision-making tests, including Wald's Sequential Probability Ratio Test (SPRT) and its variations \cite{Wald1945AMS, Lehmann1959}. In the classical two-choice SPRT test, the information accrued by a detector is processed to form a likelihood ratio; as successive samples are collected, the evolution of the likelihood ratio is equivalent to a discrete-time biased random walk \cite{BoEtAl06}.  

In continuous-time implementations of sequential binary hypothesis tests, evidence accumulation is represented through linear \cite{LaVigna1986}, or nonlinear \cite{Baras1987CDC}, stochastic differential equations. The relationship between discrete and continuous implementations of the SPRT is discussed in \cite{BoEtAl06}, where it is shown that, under the assumption of infinitesimal increments of information arriving at each moment in time, the logarithmic likelihood ratio in the SPRT converges in distribution to a stochastic differential equation with constant drift and diffusion terms: the \emph{drift-diffusion model (DDM)}. 
 
In this paper, we adopt the DDM as a basis for modeling information accumulation by a single unit, and we study a network of DDMs where there is communication of accumulating evidence among the units.  Our motivation for using the DDM stems in part from a class of models employed to formally investigate the cognitive and neural processes that underlie decisions in humans and animals \cite{Ra78}, \cite{BoEtAl06}. Notwithstanding the underlying complexity of human and animal decision making, carefully controlled decision-making experiments modeled using drift-diffusion processes have  proved instrumental in explaining the fundamental tradeoffs between speed and accuracy of a decision, and the conditions under which optimality is achieved \cite{BoEtAl06}.

Another source of motivation for focusing on networks of DDMs is their relevance to the design of multi-agent systems \cite{TLC2010}, and to the study of collective dynamics in biological systems \cite{BL2002}. Multidimensional DDMs can be interpreted as stochastic extensions of deterministic linear consensus dynamics \cite{Jadbabaie2003, OlMu04, Moreau2005}, and have been applied in \cite{YoScLe10, Med11} to analyze the performance of consensus protocols in the presence of noise. In particular, \cite{YoScLe10, Med11} investigate the robustness of consensus to communication noise through the $H_2$ norm of a reduced-order system that measures the expected steady-state dispersion of the agents around the consensus subspace.  

Rather than analyzing the collective effect of noise as in \cite{YoScLe10, Med11}, this paper focuses on assessing the contribution of each unit to the uncertainty of the process by identifying the structural elements of the network that govern the unit's individual behavior. From this perspective, our work complements research on controllability of networks \cite{Liu2011Nature}, in which the notion of nodal degree is identified as the structural element of the network that determines the units whose direct control ensures controllability of the network. The degree distribution is also important in specifying which nodes are critical to maintaining a network's structural integrity under random failures or targeted attacks \cite{Albert2000Nature}, \cite{Jeong2001Nature}. However, as we show in this work, the notion of nodal degree cannot be used to assess the significance of each unit based on certainty.

Quantifying the effect of the network structure on individual behavior naturally leads to the concept of \emph{centrality} \cite{Ne10}. Centrality measures typically assign to each node a quantity that reflects its location in the network and summarizes the node's involvement in the cohesiveness of the network process. Motivated by the structural properties of the star graph, Freeman \cite{FREE78} proposed a categorization of centrality measures based on node degree, closeness and betweenness. The degree centrality of a node depends on the number of nodes adjacent to it, and it is an intrinsically local metric of centrality. In contrast, closeness and betweenness adopt a more global perspective by relying on shortest---i.e., geodesic---paths connecting pairs of nodes \cite{FREE78}. Centrality indices based on geodesic paths are also discussed in \cite{Wuchty2003JTB}, where it is pointed out that such measures may capture certain network organizational effects that cannot be distinguished by local measures, such as degree distributions.

However, the choice of the geodesic path as the structural ingredient of the graph based on which centrality is defined cannot capture the finest structure of the network because it neglects communication along non-geodesic pathways \cite{StZe89}, \cite{Poulin2000SN}. Indeed, evidence transmitted by each unit through the network can reach the rest of the units via circuitous, not necessarily geodesic, pathways. 

To incorporate non-geodesic paths, Stephenson and Zelen proposed \emph{information centrality} and applied it to interpret the spread of an infectious disease in a network of interconnected individuals \cite{StZe89}. Following \cite{StZe89}, the work in \cite{Alt93} showed that information centrality can be reinterpreted as the electrical conductance of an equivalent electrical network and proposed an improved algorithm for its computation. This analogy reveals the intrinsic connection between information centrality and effective resistance, which was found relevant in many fields beyond electrical network analysis; see \cite{GhBoSa08} for applications including Markov chains, averaging networks, and linear parameter estimation. The connection between the estimation error covariance and the effective resistance is discussed in \cite{Barooah2007CSM} in the context of estimating vector-valued quantities from relative noisy measurements. In a different context, the total effective resistance has been related to the robustness of consensus to white noise in \cite{YoScLe10}. 

In the present work, we elucidate the relationship between the location of a node in the communication topology and its certainty as it collects and exchanges information with its neighbors; that is, the nodes with which it can communicate. We start by adopting the drift-diffusion model for sequential evidence accumulation in continuous time for an individual unit, and extending it to consider multiple such units interconnected according to communication topologies with normal graphs. 
Based on this model we introduce an index that characterizes the certainty of each unit by considering the difference between its variance and the minimum possible variance it can attain. We then provide a formal connection between the certainty index and the notion of information centrality \cite{StZe89} by proving that ordering of nodes with respect to information centrality predicts ordering of nodes with respect to certainty.  This demonstrates that collective evidence accumulation is a total network process: the entirety of paths connecting a unit with the rest of the network---including paths that are not geodesic---affect the unit's certainty as it integrates noisy information about an external signal or decision alternative. 

The structure of this paper is as follows. Section \ref{sec:TAC10:models} motivates the DDM as a model for evidence accumulation in decision making and it extends it to a network setting. Section \ref{sec:TAC10:normal_graph} defines an index for classifying the nodes according to their certainty. Section \ref{sec:TAC10:main_result} provides our main result that interprets the node certainty index based on the structural properties of the communication graph through the notion of information centrality. Sections \ref {sec:TAC10:examples_normal} and \ref{sec:TAC10:non_normal} present comparisons between certain classes of graphs in terms of the certainty of their nodes. Section \ref{sec:TAC10:conclusions} concludes the paper.

\section{Model}
\label{sec:TAC10:models}

This section describes the drift-diffusion model (DDM) of evidence accumulation and its extension to a network setting.  

\subsection{Sequential Evidence Accumulation}
\label{subsec:TAC10:DDM}

In its standard form, the DDM corresponds to the stochastic differential equation 
\begin{equation} \label{eq:DDM}
        dx =  \beta dt + \sigma dW,
\end{equation}
where $\beta$ is a constant drift term and $\sigma dW$ are increments drawn from a Wiener process with standard deviation $\sigma$. The interpretation of the random process $\{x(t):t \geq 0 \}$ evolving according to \eqref{eq:DDM} in the context of evidence accumulation is the subject of this section. 

In the decision-making literature, the DDM arises in a variety of ways; detailed accounts can be found in extensive reviews on the diffusive paradigm of decision making as in \cite{BoEtAl06} for example. One way to interpret \eqref{eq:DDM} is to consider the continuous-time limit of the logarithmic likelihood ratio in the classical SPRT. Following Wald's treatment \cite{Wald1945AMS}, suppose that $Y$ is a random variable and let $H_0$ and $H_1$ denote the hypotheses that the probability distribution of $Y$ is $p_0(y)$ (null hypothesis) and $p_1(y)$ (alternative hypothesis), respectively. The objective is to decide which hypothesis is the correct one on the basis of a sequence of independent observations $y_1, y_2,..., y_N$ of $Y$. In the SPRT, incoming data are processed to form a likelihood ratio
\begin{equation}\label{eq:LR}
\Lambda_N = \frac{p_0(y_1) p_0(y_2) \cdots p_0(y_N)}{p_1(y_1) p_1(y_2) \cdots p_1(y_N)},
\end{equation}
which summarizes the information available up to and including the current ($N$-th) observation $y_N$. If $y_N$ supports hypothesis $H_0$, the ratio $p_0(y_N) / p_1(y_N)$ is greater than one---that is, $y_N$ is more likely under $H_0$ than under $H_1$---and then $\Lambda_N$ increases. On the other hand, if $y_N$ supports $H_1$, the ratio $p_0(y_N) / p_1(y_N)$ is lower than one and $\Lambda_N$ decreases. 

Applying logarithms, the likelihood ratio \eqref{eq:LR} represents a discrete random walk evolving according to
\begin{equation}\label{eq:LOGLR}
\tilde{\Lambda}_N = \tilde{\Lambda}_{N-1} + \log \frac{p_0(y_N)}{p_1(y_N)},
\end{equation}
where $\tilde{\Lambda}_N = \log \Lambda_N$ and $\log \frac{p_0(y_N)}{p_1(y_N)}$ corresponds to the increment of information gained from observation $y_N$. It is shown in \cite{BoEtAl06} that under the assumption of infinitesimal increments of information arriving at each moment in time, and up to an unimportant scaling factor, the discrete random walk \eqref{eq:LOGLR} converges in distribution to the process described by \eqref{eq:DDM}. In light of this result, the meaning of \eqref{eq:DDM} becomes clear: its solution $\{ x(t): t \geq 0 \}$ denotes the accumulated value at time $t$ of the difference in the information favoring one hypothesis over the other, while the constant drift $\beta$ represents increase in the evidence supporting the correct decision.

In a different, neurally-motivated, decision-making context, the DDM \eqref{eq:DDM} appears as a model for evidence accumulation through appropriate reductions in models of \emph{competing leaky accumulators} as detailed in \cite{BoEtAl06}. Such models have been proposed to explain the neural mechanisms of integration of information in perceptual choice tasks \cite{USMC01}, and they correspond to two mutually inhibitory, competing neural populations, which  provide evidence supporting each of the two hypotheses. 

Other discrete- and continuous-time models that have been proposed to investigate the neural mechanisms of decision making also reduce to the DDM as a model for evidence accumulation in continuous time; see \cite{BoEtAl06} for a detailed review. The majority of these models aim to capture the phenomenology of such processes and are carefully justified through experimental data.  On the other hand, \cite{Smith10JMP} arrives at the DDM \eqref{eq:DDM} via a mechanistic approach, providing a quantitative link from the microscopic, short-time statistics of neuronal representations to the macroscopic, long-time statistics of information accumulation processes.

\subsection{Networks of Interconnected DDMs}
\label{subsec:TAC10:model}

We model a network of $n$ evidence accumulating units as the interconnection of $n$ DDMs that share the relative value of their evidence with those units with which they can communicate. In more detail, the state $x_k$ of unit $k$, for each $k = 1, \ldots, n$, evolves according to
\begin{equation} \label{eq:SDE_component}
        dx_k = \left[ \beta + \sum^n_{j = 1} \alpha_{kj} (x_j - x_k) \right] dt + \sigma dW_k,
\end{equation}
where, in analogy with \eqref{eq:DDM}, $\beta$ represents a constant drift term and $\sigma dW_k$ corresponds to increments drawn from independent Wiener processes with standard deviation $\sigma$. In \eqref{eq:SDE_component}, $\alpha_{kj} \geq 0$ denotes the attention paid by unit $k$ to the difference between its state $x_k$ and the state $x_j$  of unit $j$; $\alpha_{kj} = 0$ implies that the units $k$ and $j$ do not communicate.

The model \eqref{eq:SDE_component} can be associated with a collective decision-making scenario, in which a set of interconnected decision-making units is presented with partial information about a stimulus---e.g., a deterministic signal corrupted by noise---and each unit is asked to  identify it between two alternatives within a finite time interval \cite{PoScLe10}. 

Beyond decision making, the model \eqref{eq:SDE_component} with $\beta=0$ has been used to determine sufficient and necessary conditions for mean-square average consensus under measurement noise \cite{LIZH09}, and to analyze the stochastic stability \cite{Med11}, and robustness \cite{YoScLe10}, of linear consensus algorithms in the presence of (white) noise. A common metric for assessing the quality of consensus under measurement noise is the trace of the stationary covariance matrix associated with the projection of the state on the subspace orthogonal to the consensus subspace \cite{YoScLe10}, \cite{Med11}; see also \cite{Xiao2007JPDC} that uses a similar metric for the discrete-time case.

Such metrics capture the \emph{collective} effect of the uncertainty; but, they do not distinguish the \emph{individual} contributions of the nodes to the dispersion around the consensus subspace. It is therefore natural to ask how the uncertainty of each node affects the total uncertainty of the process, and how individual contributions can be characterized based on the locations of the nodes in the underlying interconnection graph. Our aim in this paper is to provide an answer to these questions.

\subsection{Notation and Basic Properties of the Model}
\label{subsec:properties}

It is useful to identify the communication topology in the network with a digraph ${\cal G}=({\cal V}, {\cal E}, A)$. The vertex set ${\cal V}:=\{v_1,...,v_n\}$ contains $n$ nodes that represent the $n$  evidence accumulators. The edge set ${\cal E} \subseteq {\cal V}  \times {\cal V}$ contains the communication links among the nodes, and $A \in \mathbb{R}^{n \times n}_{\geq 0}$ is the corresponding weighted adjacency matrix. The elements of $A$ are denoted by $\alpha_{kj} \geq 0$ and defined as follows: for $v_k, v_j \in {\cal V}$, $~\alpha_{kj}>0$ if $e_{kj}=(v_k,v_j) \in {\cal E}$, and $~\alpha_{kj}=0$ otherwise. Throughout this work, a ``sensing'' convention is adopted: a (directed) edge $e_{kj}=(v_k, v_j) \in {\cal E}$ implies that node $v_j$ \emph{transmits} information about its state to node $v_k$. Graphically, $e_{kj}$ is represented by an arrow from node $v_k$ to node $v_j$, implying that node $v_k$ can ``sense'' the state of node $v_j$, and we say that $v_j$ is a ``neighbor'' of $v_k$. We will assume that there are no self-loops in ${\cal G}$, i.e., $\alpha_{kk}= 0$ for all $v_k \in {\cal V}$. The out- and in-degree of a node $v_k \in {\cal V}$ can be defined by
\begin{equation} \nonumber
 {\rm {deg}}_{\rm {out}}(v_k) := \sum^n_{j=1} \alpha_{kj} \mbox{~~and~~} {\rm {deg}}_{\rm {in}}(v_k) := \sum^n_{j=1} \alpha_{jk},
\end{equation}
respectively. If ${\rm {deg}}_{\rm {out}}(v_k)={\rm {deg}}_{\rm {in}}(v_k)$ for all $v_k \in {\cal V}$, the graph ${\cal G}$ is called \emph{balanced}.

In this notation, \eqref{eq:SDE_component} takes the form
\begin{equation} \label{eq:SDE_vector}
        dx = \left( b - Lx \right) dt + H dW,
\end{equation}
where $x := {\rm {col}}(x_1,\ldots,x_n)$, $dW := {\rm {col}}(dW_{1}, \ldots,dW_n)$, $b := \beta {\bf 1}_n$ and $H := \sigma I_n$; ${\bf 1}_n$ is the $n$-dimensional vector with entries all equal to one and $I_n$ is the $n \times n$ identity matrix. In \eqref{eq:SDE_vector}, $L$ is the Laplacian matrix associated with ${\cal G}$, defined by
\begin{eqnarray}
  L_{kj} :=
  \begin{cases}
    \begin{aligned}
        \sum^n_{i=1, i \neq k} \alpha_{ki}, & ~~ k=j,\\
        -\alpha_{kj}, & ~~k \neq j.
    \end{aligned}
  \end{cases}
\end{eqnarray}
By construction, ${\bf 1}_n$ is an eigenvector of $L$ corresponding to the eigenvalue $\lambda_1 = 0$. Note that if the graph is balanced, ${\bf 1}^{\rm T}_n$ is a left eigenvector of $L$ associated with $\lambda_1 = 0$.

To fix terminology, some basic connectivity notions are now in order. A (directed) \emph{path} in a digraph ${\cal G}$ is an ordered sequence of vertices, such that any pair appearing consecutively is an edge of the digraph. A vertex of a digraph is \emph{globally reachable} if, and only if, it can be reached form any other vertex by traversing a directed path. A digraph ${\cal G}$ is \emph{strongly connected} if, and only if, every vertex is globally reachable. Next, some relevant properties of the Laplacian are summarized in Proposition \ref{prop:TAC_10:graph_Laplacian}.

\begin{proposition}\label{prop:TAC_10:graph_Laplacian}
Let ${\cal G}:=({\cal V}, {\cal E}, A)$ be a digraph of order $n$, and $L$ its associated Laplacian. Then,
\begin{enumerate}[(i)]
\item \label{thm:TAC_10:graph_Laplacian_1} all the eigenvalues of $L$ have nonnegative real parts;
\item \label{thm:TAC_10:graph_Laplacian_2} if ${\cal G}$ is strongly connected\footnote{Note that this condition can be relaxed to graphs that contain a globally reachable node. However, in Section \ref{sec:TAC10:normal_graph} we focus on strongly connected digraphs with normal Laplacian matrices, which by \cite[Lemma 4]{YoScLe10} are balanced. For balanced graphs the two conditions are equivalent; see \cite{Wu05}.}, then ${\rm {rank}}(L)=n-1$, i.e., $0$ is a simple eigenvalue of $L$;
\item \label{thm:TAC_10:graph_Laplacian_3} for any $\tau \in [0,t]$, $e^{-L(t-\tau)}$ is a row-stochastic matrix.
\end{enumerate}
\end{proposition}
Statements \eqref{thm:TAC_10:graph_Laplacian_1} and \eqref{thm:TAC_10:graph_Laplacian_2} are proved in \cite[Theorem 1.37]{BuCoMa09}, and statement \eqref{thm:TAC_10:graph_Laplacian_3} is a direct consequence of the fact that the rows of the Laplacian $L$ sum to zero.  

The following proposition characterizes the statistics of the process $\{x(t):t \geq 0 \}$ produced by \eqref{eq:SDE_vector} given deterministic zero initial conditions, i.e., ${\rm {Cov}}(x_0, x_0) = 0$ and $\mathbb{E}[x_0] = 0$. 

\begin{proposition}\label{prop:TAC_10:SDE_solution}
Let $x(0) = 0$ with probability one. Then, the general solution of \eqref{eq:SDE_vector} is
\begin{equation} \label{eq:SDE_solution}
    x(t) = { \int_0^t e^{-L(t-\tau)} b d \tau } +{\int_0^t e^{-L(t-\tau)} H dW },
\end{equation}
in which the stochastic integral is interpreted in the It\^{o} sense. In addition,
\begin{enumerate}[(i)]
\item{the mean and covariance of \eqref{eq:SDE_solution} are given by
\begin{equation} \label{eq:SDE_solution_mean}
    \mathbb{E} [x(t)] = { \int_0^t e^{-L(t-\tau)} b d\tau }
\end{equation}
and
\begin{equation} \label{eq:SDE_solution_cov}
    {\rm {Cov}} (x(t), x(t))  = \sigma^2 {\int_0^t e^{-L(t-\tau)} e^{-L^{\rm T}(t-\tau)} d\tau },
\end{equation}
respectively;}
\item{the stochastic process $\{x(t): t \geq 0 \}$ is Gaussian.}
\end{enumerate}
\end{proposition}
The proof of Proposition \ref{prop:TAC_10:SDE_solution} is a straightforward consequence of  \cite[pp. 131--132]{AR74}. 
%
The lemma below provides lower and upper bounds for the variance of the state of each unit. The lower bound will be important in defining an index that characterizes the certainty of each unit as will be discussed in Section \ref{sec:TAC10:normal_graph}.

\begin{lemma}\label{lem:TAC_10:cor_1}
Consider \eqref{eq:SDE_vector}. For any interconnection digraph ${\cal G}=({\cal V}, {\cal E}, A)$ and any node $v_k \in {\cal V}$,
\begin{equation} \label{eq:SDE_solution_mean_node}
    \mathbb{E} [x_k(t)] =  \beta t
\end{equation}
and
\begin{equation} \label{eq:SDE_solution_var_node_bound}
   \frac{\sigma^2}{n} t \leq {\rm {Var}} (x_k(t))  \leq \sigma^2 t.
\end{equation}
\end{lemma}
\begin{IEEEproof}
Equation \eqref{eq:SDE_solution_mean_node} is an immediate consequence of \eqref{eq:SDE_solution_mean} for $b = \beta {\bf 1}_n$, since, by Proposition \ref{prop:TAC_10:graph_Laplacian}\eqref{thm:TAC_10:graph_Laplacian_3}, $e^{-L(t-\tau)}$ is a row-stochastic matrix. To show \eqref{eq:SDE_solution_var_node_bound}, let ${\bf q}_k$ be the $n \times 1$ vector with all elements equal to zero except element $k$, which is equal to one; note that $\sum^n_{k=1} {\bf q}_k = {\bf 1}_n$. Then,
\begin{equation}
\begin{aligned}
	{\rm {Var}}(x_k(t)) &= {\bf q}^{\rm T}_k  {\rm {Cov}}(x(t), x(t)) {\bf q}_k \\
	&= \sigma^2 {\int_0^t ||e^{-L^{\rm T}(t-\tau)} {\bf q}_k ||^2 d\tau } \\
	&= \sigma^2 {\int_0^t \sum^n_{\ell = 1} \left({\bf q}^{\rm T}_\ell e^{-L^{\rm T}(t-\tau)} {\bf q}_k \right)^2d\tau },
\end{aligned}
\end{equation}
where \eqref{eq:SDE_solution_cov} has been used. The lower bound in \eqref{eq:SDE_solution_var_node_bound} is obtained through Jensen's inequality\footnote{Jensen's inequality: Let $f$ be a convex function on an interval ${\cal J}$ and $x_j \in {\cal J}$ for $j \in \{1, ..., n\}$. Then, $f \left( \frac{1}{n} \sum^n_{j=1} x_j \right) \leq \frac{1}{n} \left( \sum^n_{j=1} f(x_j) \right).$},
\begin{equation}
\begin{aligned}
	\sum^n_{\ell = 1} \left( {\bf q}^{\rm T}_\ell e^{-L^{\rm T}(t-\tau)} {\bf q}_k \right)^2 &\geq \frac{1}{n} \left( \sum^n_{\ell = 1}  {\bf q}^{\rm T}_\ell e^{-L^{\rm T}(t-\tau)} {\bf q}_k \right)^2,
\end{aligned}
\end{equation}
by observing that $e^{-L^{\rm T}(t-\tau)} {\bf q}_k$  corresponds to the $k$-th column of $e^{-L^{\rm T}(t-\tau)}$, that is, the $k$-th row of $e^{-L(t-\tau)}$, and by noticing that $e^{-L(t-\tau)}$ is row-stochastic by Proposition \ref{prop:TAC_10:graph_Laplacian}\eqref{thm:TAC_10:graph_Laplacian_3}. Finally, the upper bound  follows from
\begin{equation}
\begin{aligned}
\sum^n_{\ell = 1} \left( {\bf q}^{\rm T}_\ell e^{-L^{\rm T}(t-\tau)} {\bf q}_k \right)^2 &\leq \left( \sum^n_{\ell = 1} \left| {\bf q}^{\rm T}_\ell e^{-L^{\rm T}(t-\tau)} {\bf q}_k \right| \right)^2
\end{aligned}
\end{equation}
in a similar fashion.
\end{IEEEproof}

\begin{rem}\label{rem:interconnection_cov}
%
%
Lemma \ref{lem:TAC_10:cor_1} shows that the expected value of the evidence accumulated by each unit increases linearly with time at a rate $\beta$, which is the same for all units regardless of the interconnection topology. By way of contrast, the covariance matrix \emph{does} depend on the interconnection. This implies that certain communication topologies---and certain nodes within them---may be better than others in terms of certainty in integrating information. In view of the fact that $\sigma^2 t$ is the variance of the state of an isolated DDM, the upper bound in \eqref{eq:SDE_solution_var_node_bound} implies that the uncertainty associated with any of the interconnected units cannot exceed that of an isolated unit.
%
\end{rem}
\begin{rem}
When $t$ is sufficiently small, by expanding the exponentials in \eqref{eq:SDE_solution_cov} in Taylor series and neglecting higher order terms, ${\rm {Cov}} ( x(t),x(t) ) \approx (\sigma^2 t) I_n$. This fact implies that all units behave like isolated DDMs at the beginning of the process. It will become apparent in the following sections that, as time evolves and the units collect and communicate their accumulated evidence, their certainty improves with respect to that of an isolated DDM in a way that depends on the topology of the communication. We are interested to identify units with variance that evolves more closely to the lower bound in \eqref{eq:SDE_solution_var_node_bound}.
\end{rem}

\section{Node Certainty}
\label{sec:TAC10:normal_graph}

This section introduces an index that characterizes the certainty of each unit as it accrues evidence. We will restrict our analysis to strongly connected digraphs ${\cal G}=({\cal V}, {\cal E}, A)$ with Laplacian matrices $L$ that are normal; i.e., matrices that commute with their transpose, \cite[Sec. 2.5]{HoJo85}. 


For each $v_k \in {\cal V}$ we define the \emph{node certainty index} $\mu : {\cal V} \to \mathbb{R}_{> 0} \cup \{\infty\}$ as the inverse of the difference between the variance ${\rm {Var}}(x_k(t))$ of the state $x_k$ of node $v_k$ and the minimum possible variance $\sigma^2 t / n$ as $t \to +\infty$; that is,
\begin{equation} \label{eq:TAC10:mu_definition}
\frac{1}{\mu(v_k)} := \lim_{t \to +\infty} \left( {\rm {Var}}(x_k(t)) -  \sigma^2 \frac{t}{n} \right).
\end{equation}
A high value of $\mu(v_k)$ corresponds to small uncertainty associated with the node $v_k$, since the variance of its state evolves closely to the minimum possible variance $\sigma^2 t / n$; see Lemma \ref{lem:TAC_10:cor_1}. By convention, $\mu(v_k)=\infty$ corresponds to the highest possible certainty. 

Before we continue with interpreting the index \eqref{eq:TAC10:mu_definition} based on properties of the interconnection graph, the following proposition shows that $\mu$ is well defined and provides a formula for the computation of $\mu$ in terms of the eigenstructure of the Laplacian. 

\begin{proposition}
\label{prop:TAC_10:node_certainty_index}
Let ${\cal G}:=({\cal V}, {\cal E}, A)$ be a digraph, and $L$ its associated Laplacian. Assume that ${\cal G}$ is strongly connected and that $L$ is normal. Then,
\begin{enumerate}[(i)]
\item \label{thm:TAC_10:certainty_index_definition} the limit in \eqref{eq:TAC10:mu_definition} is well defined;
\item \label{thm:TAC_10:certainty_index_computation} the index $\mu$ can be computed by
\begin{equation}\label{eq:TAC10:mu_computation}
\frac{1}{\mu(v_k)} = \sigma^2 \sum^{n}_{p=2} \frac{1}{2 {\rm {Re}}(\lambda_p)} \big| u^{(p)}_k \big|^2, 
\end{equation}
where ${\rm {Re}}(\lambda_p)$ denotes the real part of the nonzero eigenvalue $\lambda_p$, $p \in \{2,...,n\}$ of $L$, and $u^{(p)}_k$ is the $k$-th component of the $p$-th normalized eigenvector.
\end{enumerate}
\end{proposition}
To prove Proposition \ref{prop:TAC_10:node_certainty_index}, we will use the following lemma that provides an analytical expression for the covariance matrix \eqref{eq:SDE_solution_cov}.

\begin{lemma}\label{lem:normal_var}
Consider \eqref{eq:SDE_vector}. Under the conditions and notation of Proposition \ref{prop:TAC_10:node_certainty_index}, the elements of the covariance matrix are
\begin{equation} \label{eq:cov_normal_entries}
    [{\rm {Cov}}(x(t),x(t))]_{k j} =  \sigma^2 \frac{t}{n} + \sigma^2 \sum^{n}_{p=2} \frac{1 - e^{-2 {\rm {Re}}(\lambda_p) t}}{2 {\rm {Re}}(\lambda_p)} u^{(p)}_k \bar{u}^{(p)}_j,
\end{equation}
where $\bar{u}^{(p)}_k$ denotes the complex conjugate of $u^{(p)}_k$.
\end{lemma}
%
\begin{IEEEproof}
By the normality of $L$ there exists a unitary matrix $U$, such that $U^* L U = \Lambda$, where $U^*$ is the Hermitian transpose of $U$ and $\Lambda$ is a diagonal matrix containing the eigenvalues of $L$. Substitution in \eqref{eq:SDE_solution_cov} results in
\begin{equation} \label{eq:SDE_solution_cov_normal}
         {\rm {Cov}} (x(t), x(t)) = \sigma^2~\left( U~G(t)~U^* \right),
\end{equation}
where
\begin{equation} \label{eq:SDE_solution_cov_normal_G}
         G(t) := {\int_0^t e^{-\left(\Lambda + \bar{\Lambda}\right) (t-\tau)} d\tau },
\end{equation}
and $\bar{\Lambda}$ is the diagonal matrix containing the complex conjugates of the eigenvalues of $L$. Letting $U = [u^{(1)}|\ldots|u^{(n)}]$, \eqref{eq:SDE_solution_cov_normal} gives
\begin{equation}
    \left[{\rm {Cov}}(x(t), x(t))\right]_{k j} = \sigma^2 \sum^{n}_{p=1} g_{p p}(t) u^{(p)}_k \bar{u}^{(p)}_j,
\end{equation}
in which $g_{p p}(t)$ denotes the $p$-th element of the diagonal matrix $G(t)$ and is computed by \eqref{eq:SDE_solution_cov_normal_G} as
\begin{eqnarray}
  g_{p p}(t) :=
  \begin{cases}
    \begin{aligned}
        t, & ~~ \mbox{if~} \lambda_p = 0,\\
        \frac{1 - e^{-2 {\rm {Re}}(\lambda_p) t}}{2 {\rm {Re}}(\lambda_p)}, & ~~ \mbox{if~} \lambda_p \neq 0.
    \end{aligned}
  \end{cases}
\end{eqnarray}
Since the graph is assumed to be strongly connected, by Proposition \ref{prop:TAC_10:graph_Laplacian}\eqref{thm:TAC_10:graph_Laplacian_2} ${\bf 1}_n$ spans the kernel of $L$, implying that $u^{(1)} = \left(1 / \sqrt{n} \right){\bf 1}_n$ is the normalized eigenvector corresponding to the zero eigenvalue $\lambda_1$ of $L$, thus resulting in \eqref{eq:cov_normal_entries}.
\end{IEEEproof}

With the aid of Lemma \ref{lem:normal_var} we can now proceed with a proof of Proposition \ref{prop:TAC_10:node_certainty_index}.

\begin{IEEEproof}
From Lemma \ref{lem:normal_var} 
\begin{equation} \label{eq:var_normal}
    {\rm {Var}}(x_k(t)) =  \sigma^2 \frac{t}{n} + \sigma^2 \sum^{n}_{p=2} \frac{1 - e^{-2 {\rm {Re}}(\lambda_p) t}}{2 {\rm {Re}}(\lambda_p)}~ \big| u^{(p)}_k \big|^2.
\end{equation}
By Proposition \ref{prop:TAC_10:graph_Laplacian}, strong connectivity of ${\cal G}$ implies ${\rm {Re}}(\lambda_p) > 0$ for $p=2,...,n$, and \eqref{eq:var_normal} indicates that 
\begin{equation}
\label{eq:var_normal_inequality}
{\rm {Var}}(x_k(t)) -  \sigma^2 \frac{t}{n} ~\leq~ \sigma^2 \sum^{n}_{p=2} \frac{1}{2 {\rm {Re}}(\lambda_p)} \big| u^{(p)}_k \big|^2,
\end{equation}
uniformly in $t$. In \eqref{eq:var_normal_inequality} equality is asymptotically attained as $t \to +\infty$, proving both parts of Proposition \ref{prop:TAC_10:node_certainty_index}.
\end{IEEEproof}

The dependence of the node accuracy index $\mu$ on the eigenstructure of the graph Laplacian $L$ according to \eqref{eq:TAC10:mu_computation} reflects the fact that the certainty of each node is contingent upon its location in the underlying interconnection graph. Classifying the nodes of a graph based on their certainty and interpreting this classification in terms of the structural properties of the interconnection graph will be discussed in Section \ref{sec:TAC10:main_result} below. The following remark provides further intuition about the index $\mu$.

\begin{rem}\label{rem:H2_norm}
From \eqref{eq:TAC10:mu_computation} it is easy to see that
\begin{equation}\label{eq:sum_accuracy_index}
\sum_{v_k \in {\cal V}({\cal G})} \frac{1}{\mu(v_k)} = \sigma^2 \sum^{n}_{p=2} \frac{1}{2 {\rm {Re}}(\lambda_p)}.
\end{equation}
As was discussed in \cite{YoScLe10} in the context of linear consensus protocols in the presence of additive white noise, the sum in the right hand side of \eqref{eq:sum_accuracy_index} corresponds to the expected steady-state \emph{dispersion} around the consensus subspace. 
%
Hence, the inverse of $\mu(v_k)$ can be interpreted as the \emph{individual} contribution of the node $v_k$ to the dispersion of the evidence; the higher $\mu(v_k)$, the smaller the contribution of the node $v_k$. In the case of undirected graphs, the sum \eqref{eq:sum_accuracy_index} is related to the effective resistance $K_{\rm f}$, or Kirchhoff index, of the graph 
\begin{equation}\label{eq:kirchhoff_index}
K_{\rm f} := n \sum^n_{p=2} \frac{1}{\lambda_p};
\end{equation}
see \cite{GhBoSa08}. Clearly, for undirected graphs
\begin{equation}\label{eq:kirchhoff_index_mu}
\sum_{v_k \in {\cal V}({\cal G})} \frac{1}{\mu(v_ k)} = \sigma^2 \left(\frac{K_{\rm f}}{2n}\right).
\end{equation}
\end{rem}
\section{Node Certainty as a Centrality Measure}
\label{sec:TAC10:main_result}
 
In this section, the node certainty index $\mu$ is characterized in terms of the structural properties of the underlying interconnection graph. It is intuitively discussed in Section \ref{subsec:TAC10:SZ_example} that node certainty depends on the \emph{totality} of paths---and not just the geodesic paths---in the network. This observation is rigorously formalized in Section \ref{subsec:TAC10:main_result}, which reinterprets the node certainty index as a centrality measure by establishing its connection with the notion of information centrality.

\subsection{Motivation}
\label{subsec:TAC10:SZ_example}

\begin{figure}[t!]
{\centering
\begin{minipage}{0.5\columnwidth}    
  	\centering
  	{\includegraphics[width=0.3\textwidth]{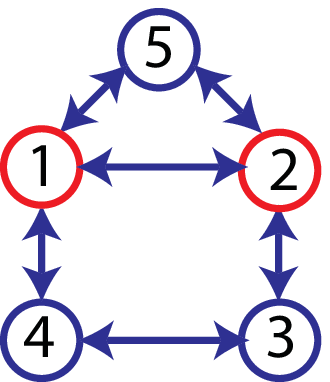}}
	\label{fig:star_graph_expl}
\end{minipage}
\begin{minipage}{0.5\columnwidth}     
	\centering
  	\begin{tabular}{c c c c}
	\hline
	\bf{Vertex}  & \bf{Degree} & $\kappa_{\rm {close}}$  & $\mu$ \\ 
	\hline
	$v_1$ & $3$ & $1.00$ & $8.33$ \\ 
	$v_2$ & $3$ & $1.00$ & $8.33$ \\ 
	$v_3$ & $2$ & $0.83$ & $5.26$ \\ 
	$v_4$ & $2$ & $0.83$ & $5.26$ \\ 
	$v_5$ & $2$ & $0.83$ & $5.00$ \\ 
	\hline
	\end{tabular}
\end{minipage}
}
\caption{Left: The connected undirected graph used to illustrate that \emph{all} paths---not just the geodesic ones---must be taken into account in interpreting $\mu$. The nodes that maximize the certainty index are $v_1$ and $v_2$, and the node that minimizes certainty is  $v_5$. Right: The table summarizes properties of the nodes; namely, degree, $\kappa_{\rm {close}}$ and $\mu$.}
\label{fig:s_z_example}
\end{figure}

To motivate the discussion, we provide an example of an undirected graph; see Fig. \ref{fig:s_z_example}. For each node $v_k$, we compute the certainty index $\mu(v_k)$ using \eqref{eq:TAC10:mu_computation}, and provide its degree; that is, the number of edges attached to $v_k$. In addition, the corresponding \emph{closeness} centrality, $\kappa_{\rm {close}}$, is provided as a representative geodesic-distance-based measure of centrality. Defining the geodesic distance $d(v_k,v_j)$ between $v_k$ and $v_j$ as the length of the \emph{shortest} path connecting them, the closeness centrality of a node $v_k$ is computed as the inverse of the mean geodesic distance $d(v_k,v_j)$ averaged over all nodes $v_j$, i.e.
\begin{equation}\label{eq:closeness_centrality}
\kappa_{\rm {close}}(v_k) = \left( \frac{1}{n} \sum^n_{j=1} d(v_k,v_j) \right)^{-1};
\end{equation}
see \cite[Section 7.6]{Ne10}. The example of the graph of Fig. \ref{fig:s_z_example} demonstrates that, for general undirected graphs, node certainty cannot be captured by centrality measures based on degrees or geodesic paths. This is a consequence of the fact that the evidence accumulated by each unit is transmitted through the network and reaches the rest of the units via circuitous, non-geodesic pathways.  
 
In more detail, from Fig. \ref{fig:s_z_example} note that $\mu(v_3) = \mu(v_4) > \mu(v_5)$. This distinction between $v_5$ and $\{v_3, v_4\}$ cannot be captured by their degrees, which are all equal to $2$. 
%
Closeness centrality too cannot discriminate between $v_4$ and $v_5$. In fact, the different certainty levels of $v_4$ and $v_5$ cannot be captured by \emph{any} centrality measure that is defined based on geodesic paths. To see this, note that any of the vertices $v_3$, $v_4$ and $v_5$ in the graph of Fig. \ref{fig:s_z_example} is connected to the rest through two geodesic paths of length $2$ and two geodesic paths of length $1$. Hence, excluding non-geodesic pathways, these nodes are equivalent, resulting in $\kappa_{\rm {close}}(v_3)=\kappa_{\rm {close}}(v_4)=\kappa_{\rm {close}}(v_5)$.

To provide further intuition, consider the pairs $\{v_1, v_4\}$ and $\{v_1, v_5\}$, and enumerate all possible paths connecting them. For $\{v_1, v_4\}$ we have the paths $v_1-v_4$, $v_1-v_2-v_3-v_4$ and $v_1-v_5-v_2-v_3-v_4$; for $\{v_1,v_5\}$ we have the paths $v_1-v_5$, $v_1-v_2-v_5$ and $v_1-v_4-v_3-v_2-v_5$. Thus, the evidence transmitted by $v_1$ reaches $v_4$ via three paths of length $1$, $3$ and $4$, respectively. On the other hand, it reaches $v_5$ via three paths of lengths $1$, $2$ and $4$, respectively. This difference is reflected in the node certainty index, revealing the non-geodesic nature underlying information transmission, which requires that \emph{all} possible paths between any pair of nodes in the network must be taken into account.

\subsection{Main Result: Node Certainty and Information Centrality}
\label{subsec:TAC10:main_result}

This section clarifies the relation between node certainty, as characterized by the index $\mu$, and the location of a node in the underlying interconnection graph through the notion of \emph{information centrality} \cite{StZe89}. 

To define information centrality, we begin with a weighted undirected graph $\hat{\cal G}=(\hat{\cal V}, \hat{\cal E}, \hat{A})$, which is assumed to be connected. Let $w: \hat{\cal E} \to \mathbb{R}_{> 0}$ be a function that assigns to each edge $e \in \hat{\cal E}$ a positive weight $w(e)$ and consider a pair of vertices $v_k, v_j \in \hat{\cal V}$. Suppose there are $m_{kj}$ paths ${\cal P}_{kj}(r)$, $r = 1,...,m_{kj}$, connecting $v_k$ and $v_j$ and define the \emph{weighted length} of path ${\cal P}_{kj}(r)$ as
\begin{equation}\label{eq:TAC10:inverse_weighted_length}
\ell_w({\cal P}_{kj}(r)) := \sum_{e \in {\cal P}_{kj}(r)} \frac{1}{w(e)}.
\end{equation}
The definition of the length $\ell_w$ in \eqref{eq:TAC10:inverse_weighted_length} reflects the convention that the higher the weight of an edge the more important the communication between the incident nodes of that edge is; hence, these nodes appear to be ``closer''. 

To capture the effect of non-geodesic pathways, we define the distance between two nodes $v_k$ and $v_j$ based on a ``combined'' path $\tilde{\cal P}_{kj}$ that incorporates all the paths ${\cal P}_{kj}(r)$, $r = 1,...,m_{kj}$, connecting $v_k$ and $v_j$. To do so, define the $m_{kj} \times m_{kj}$ matrix $D_{kj}$ as follows:  its diagonal entries $D_{kj}(r,r)$ correspond to the weighted lengths of the paths ${\cal P}_{kj}(r)$,
\begin{equation}
D_{kj}(r,r) = \ell_w({\cal P}_{kj}(r)),
\end{equation}
and its off-diagonal entries $D_{kj}(r,s)$ correspond to the sum of the inverse weights of the edges that are common between paths ${\cal P}_{kj}(r)$ and ${\cal P}_{kj}(s)$ for $r, s \in \{1,...,m_{kj}\}$ with $r \neq s$, i.e.
\begin{equation}
D_{kj}(r,s) = \sum_{e \in {\cal P}_{kj}(r) \cap {\cal P}_{kj}(s)} \frac{1}{w(e)}.
\end{equation}
Then, the length $\ell_w(\tilde{\cal P}_{kj})$ of the combined path is given by
\begin{equation}\label{eq:combined_path_length}
\frac{1}{\ell_w(\tilde{\cal P}_{kj})} = \sum^{m_{kj}}_{r=1} \sum^{m_{kj}}_{s=1} D^{-1}_{kj}(r,s),
\end{equation}
and the distance between $v_k$ and $v_j$ is defined as
\begin{equation}\label{eq:TAC10:distance_combined_path}
\tilde{d}(v_k,v_j) := \ell_w(\tilde{\cal P}_{kj}).
\end{equation}
%
%
Stephenson and Zelen in \cite{StZe89} define the total ``information'' contained in the entirety of paths connecting $v_k$ and $v_j$ as the inverse of the length of the combined path
\begin{equation}\label{eq:combined_path_info}
I_{kj} := \frac{1}{\ell_w(\tilde{\cal P}_{kj})},
\end{equation}
with $\ell_w(\tilde{\cal P}_{kj})$ computed by \eqref{eq:combined_path_length}, and use $I_{kj}$ to compute information centrality of a node $v_k$ as the harmonic average\footnote{An alternative definition was provided in \cite{Alt93} that uses the arithmetic instead of the harmonic average. Note that the two versions of information centrality do not always give the same ranking; for instance, in the example of Fig. \ref{fig:s_z_example} node $v_5$ ranks lower than $v_4$ with the classical definition, but it ranks higher according to the definition with the arithmetic average.}
\begin{equation}\label{eq:info_centrality}
\kappa_{\rm {info}}(v_k) := \left( \frac{1}{n} \sum^n_{j=1} \frac{1}{I_{kj}} \right)^{-1} = \left( \frac{1}{n} \sum^n_{j=1} \tilde{d}(v_k,v_j) \right)^{-1}.
\end{equation}

The connection between information centrality and node certainty can now be established. Theorem \ref{thm:TAC10:information_centrality_ranking} below relates the certainty index of a node $v_k$ in a strongly connected weighted digraph ${\cal G} = ({\cal V}, {\cal E}, A)$ with normal Laplacian to the information centrality of $v_k$ in the mirror graph $\hat{\cal G}$ of ${\cal G}$. To define the mirror graph $\hat{\cal G}$ of ${\cal G}$ suppose that $\tilde{\cal E}$ is the set of reverse edges of ${\cal G}$, obtained by reversing the order of nodes of all pairs in ${\cal E}$. The mirror $\hat{\cal G}$ of ${\cal G}$ is an undirected graph $\hat{\cal G} = (\hat{\cal V}, \hat{\cal E}, \hat A)$ with set of vertices $\hat{\cal V}={\cal V}$, set of edges $\hat{\cal E} := {\cal E} \cup \tilde{\cal E}$, and with adjacency matrix $\hat A = [\hat{\alpha}_{kj}]$ with entries
\begin{equation}\nonumber
\hat \alpha_{kj} = \hat \alpha_{jk} = \frac{\alpha_{kj}+\alpha_{jk}}{2};
\end{equation}
see also \cite[Def. 2]{OlMu04}.

With this notation, the following result can be stated.

\begin{theorem}\label{thm:TAC10:information_centrality_ranking}
Let ${\cal G}=({\cal V}, {\cal E}, A)$ be a strongly connected digraph on $n$ vertices and assume that its Laplacian matrix $L$ is normal. Then, the certainty index of the node $v_k \in {\cal V}$ is 
\begin{equation}\label{eq:mu_info_graph}
\frac{1}{\mu(v_k)} = \frac{\sigma^2}{2} \left( \frac{1}{\hat \kappa_{\rm {info}}(v_k)} - \frac{\hat K_{\rm f}}{n^2} \right)
\end{equation}
where $\hat \kappa_{\rm {info}}(v_k)$ is the information centrality of $v_k$ in the mirror graph $\hat{\cal G}$ of ${\cal G}$ and $\hat K_{\rm f}$ is the Kirchhoff index of $\hat{\cal G}$ given by \eqref{eq:kirchhoff_index}. Hence, if $k_1, k_2, ..., k_n$ are indices such that
\begin{equation}
\mu(v_{k_1}) \geq \mu(v_{k_2}) \geq ... \geq \mu(v_{k_n}),
\end{equation}
then,
\begin{equation}
\hat \kappa_{\rm {info}}(v_{k_1}) \geq \hat \kappa_{\rm {info}}(v_{k_2}) \geq ... \geq \hat \kappa_{\rm {info}}(v_{k_n}),
\end{equation}
and vice versa.
\end{theorem}

Before continuing with the proof of Theorem \ref{thm:TAC10:information_centrality_ranking}, which is the subject of Section \ref{subsec:TAC10:proof} below, a remark is in order.


\begin{rem}\label{rem:trees_centrality}
In the case where the graph ${\cal G}$ is an undirected tree, for every pair of nodes $v_k, v_j \in {\cal G}$ there exists a unique path ${\cal P}_{kj}$ connecting them. Then, \eqref{eq:combined_path_info} implies that the total information transmitted between $v_k$ and $v_j$ is equal to the inverse of the weighted length $\ell_w({\cal P}_{kj})$ of ${\cal P}_{kj}$. Hence, \eqref{eq:info_centrality} reduces to \eqref{eq:closeness_centrality}, which indicates that closeness centrality can be used to discriminate the nodes of undirected trees.
\end{rem}

\subsection{Proof of Main Result}
\label{subsec:TAC10:proof}
 
In this section, Theorem \ref{thm:TAC10:information_centrality_ranking} is proved through a sequence of lemmas. 
We start with a lemma due to Stephenson and Zelen \cite{StZe89}, which provides a way to compute $I_{kj}$ defined by \eqref{eq:combined_path_info} without path enumeration.  

\begin{lemma}[Stephenson and Zelen, \cite{StZe89}]\label{lem:TAC10:stephenson_zellen}
Let $\hat{\cal G} \!=\! (\hat{\cal V}, \hat{\cal E}, \hat A)$ be an undirected connected graph of order $n$ and let $\hat L$ be its Laplacian. Then, the total information $I_{kj}$ transmitted via all paths connecting $v_k, v_j \in \hat{\cal V}$ is 
\begin{equation}\label{eq:combined_path_info_matrix}
I_{kj} = \left( c_{kk} + c_{jj} - 2 c_{kj} \right)^{-1},
\end{equation}
where $c_{kj}$, $1 \leq k,j \leq n$, are the entries of the matrix 
\begin{equation}\label{eq:s_z_matrix}
C = \left( \hat L +  {\bf 1}_n{\bf 1}^{\rm T}_n \right)^{-1}.
\end{equation}
\end{lemma}

Next, we provide a lemma that relates the node certainty index $\mu$ to the group inverse of the Laplacian $\hat{L}$ of the mirror graph $\hat{\cal G}$ of the interconnection digraph ${\cal G}$ with normal Laplacian $L$. Recall that the group inverse of an $n \times n$ matrix $P$, when it exists, is the unique matrix $X$ that satisfies:
\begin{equation}\label{eq:group_inverse_definition}
\mbox{(i)~} P X P = P, ~\mbox{(ii)~} X P X = X, ~\mbox{and (iii)~} P X = X P;
\end{equation}
see \cite[Sec. 4.4]{BeGr03} for details. In what follows, the group inverse of a matrix $P$ is denoted by $P^\#$. 

\begin{lemma}\label{lem:laplacian_group_inverse}
Let ${\cal G}$ be a strongly connected digraph with normal Laplacian matrix $L$.
\begin{enumerate}[(i)]
\item \label{lem:laplacian_group_inverse_1} The symmetric part of $L$,
\begin{equation}\label{eq:laplacian_mirror}
\hat{L} := \frac{L+L^{\rm T}}{2},
\end{equation}
is a well-defined Laplacian for the mirror graph $\hat{\cal G}$ of ${\cal G}$.
\item \label{lem:laplacian_group_inverse_2} The group inverse $\hat{L}^\#$ of $\hat{L}$ exists and is unique.
\item \label{lem:laplacian_group_inverse_3} Let $U = \left[u^{(1)} ~|~ U_{\rm r}\right]$ be the unitary matrix that diagonalizes $L$, where $u^{(1)} = (1/ \sqrt{n}) {\bf 1}_n$ is the normalized eigenvector of $L$ corresponding to the zero eigenvalue $\lambda_1 = 0$ and $U_{\rm r}$ is the $n \times (n-1)$ matrix containing the rest of the normalized eigenvectors of $L$. Then,
\begin{equation}\label{eq:g_inverse_L}
\hat{L}^\# = U_{\rm r} \left( U^*_{\rm r} \hat{L} U_{\rm r} \right)^{-1} U^*_{\rm r}.
\end{equation}
Moreover, 
\begin{equation}\label{eq:g_inverse_diagonal}
\frac{1}{\mu(v_k)} = \frac{\sigma^2}{2} \hat{L}^\#_{kk}.
\end{equation}
\end{enumerate}
\end{lemma}

\begin{IEEEproof} 
(i) By \cite[Lemma 4]{YoScLe10}, strongly connected digraphs with normal Laplacians are balanced. The result follows from \cite[Theorem 7]{OlMu04}, which states that if ${\cal G}$ is balanced, the symmetric part $\hat{L}$ of its Laplacian $L$ is a valid Laplacian matrix for the mirror graph $\hat{\cal G}$ of ${\cal G}$.

(ii) By \cite[Theorem 1, p. 162]{BeGr03}, the group inverse of $\hat{L}$ exists and is unique if, and only if, ${\rm {rank}} \hat{L} = {\rm {rank}} \hat{L}^2$; that is, if, and only if, the index of $\hat{L}$ is one. Since $\hat{L}$ is singular, this condition is equivalent to requiring that the Jordan blocks corresponding to the eigenvalue $\lambda=0$ are all $1 \times 1$; see \cite[Theorem 6, p. 170]{BeGr03}. Hence, in view of Proposition \ref{prop:TAC_10:graph_Laplacian}\eqref{thm:TAC_10:graph_Laplacian_2}, the group inverse of the Laplacian $\hat{L}$ of the mirror graph of a strongly connected digraph exists and is unique. 

(iii) Since $L$ is normal and $U$ the unitary matrix that diagonalizes $L$, we have $L U = U \Lambda$, where $\Lambda$ is a diagonal matrix containing the eigenvalues of $L$. Using the property $U^{-1} = U^*$ we obtain $U^* L = \Lambda U^*$, which, by the fact that $L$ is real and $\Lambda$ diagonal, results in $L^{\rm T} U = U \bar \Lambda$, where $\bar \Lambda$ is the complex conjugate of $\Lambda$. Since $U = \left[u^{(1)} ~|~ U_{\rm r}\right]$, these observations imply
\begin{equation} \nonumber 
\left( L+L^{\rm T} \right) U_{\rm r} = U_{\rm r} \left( \Lambda_{\rm r} + \bar{\Lambda}_{\rm r} \right), 
\end{equation}
where $\Lambda_{\rm r} = U^*_{\rm r} L U_{\rm r}$. Using this fact, and the properties
\begin{eqnarray}
\label{eq:reduced_UUstar}
U_{\rm r} U^*_{\rm r} &=& I_n - \frac{1}{n} {\bf 1}_{n}{\bf 1}^{\rm T}_{n}, \\
\label{eq:reduced_UstarU}
U^*_{\rm r} U_{\rm r} &=& I_{n-1},
\end{eqnarray}
it is straightforward to show that the matrix 
\begin{equation}\nonumber
X = U_{\rm r} \left( U^*_{\rm r} \hat{L} U_{\rm r} \right)^{-1} U^*_{\rm r}
\end{equation}
satisfies the requirements \eqref{eq:group_inverse_definition} for the group inverse; hence, we deduce that $\hat L^\# = X$. Furthermore,
\begin{eqnarray}
\nonumber
\hat{L}^\# = U_{\rm r} \left( \frac{\Lambda_{\rm r} + \bar\Lambda_{\rm r}}{2} \right)^{-1} U^*_{\rm r},
\end{eqnarray}
from which we obtain that the $(k,j)$-th entry of $\hat L^\#$ is
\begin{equation} \label{eq:TAC10:group_inverse_elements}
\hat{L}^\#_{kj} = 2 \sum^n_{p=2} \frac{1}{\lambda_p + \bar\lambda_p} u^{(p)}_k \bar u^{(p)}_j.
\end{equation}
The result \eqref{eq:g_inverse_diagonal} follows for $k=j$ in view of \eqref{eq:TAC10:mu_computation}.
\end{IEEEproof}

The following lemma collects some useful properties of $\hat{L}^\#$. 

\begin{lemma}\label{lem:laplacian_group_inverse_properties}
Let $\hat{L}^\#$ be the group inverse of the Laplacian $\hat{L}$ of a connected undirected graph $\hat{\cal G}$. Then,
{\setlength\arraycolsep{0.1em}
\begin{eqnarray}
\label{eq:TAC10:laplacian_group_inverse_properties_1}
\hat{L}\hat{L}^\# &=& \hat{L}^\#\hat{L} ~= I_n - \frac{1}{n} {\bf 1}_{n}{\bf 1}^{\rm T}_{n}, \\
\label{eq:TAC10:laplacian_group_inverse_properties_2}
{\bf 1}^{\rm T}_{n} \hat{L}^\# &=& \hat{L}^\# {\bf 1}_{n} = 0, \\
\label{eq:TAC10:laplacian_group_inverse_properties_3}
{\rm {Tr}}(\hat{L}^\#) &=& \frac{\hat{K}_{\rm f}}{n},
\end{eqnarray}}
\noindent where $\hat{K}_{\rm f}$ is the Kirchhoff index of $\hat{\cal G}$.
\end{lemma}
\begin{IEEEproof}
Equations \eqref{eq:TAC10:laplacian_group_inverse_properties_1} and \eqref{eq:TAC10:laplacian_group_inverse_properties_2} follow from \eqref{eq:g_inverse_L} in view of \eqref{eq:reduced_UUstar}-\eqref{eq:reduced_UstarU} and of the facts that ${\bf 1}^{\rm T}_{n} U_{\rm r} = U^*_{\rm r} {\bf 1}_{n} = 0$. For \eqref{eq:TAC10:laplacian_group_inverse_properties_3} we note that, by the proof of Lemma \ref{lem:laplacian_group_inverse}, if $\lambda_p$ is an eigenvalue of $L$, then $\hat{\lambda}_p = (\lambda_p + \bar{\lambda}_p)/2 = {\rm {Re}}(\lambda_p)$ is an eigenvalue of $\hat{L}$. Then, by \eqref{eq:TAC10:group_inverse_elements},
\begin{equation}\nonumber
{\rm {Tr}}(\hat{L}^\#) = \sum^n_{k=1} \hat{L}^\#_{kk} = \sum^n_{k=1} \frac{1}{\hat{\lambda}_p} = \frac{\hat{K}_{\rm f}}{n}.
\end{equation}
by the definition of the Kirchhoff index \eqref{eq:kirchhoff_index}. 
\end{IEEEproof}

The following lemma establishes a correspondence between the group inverse $\hat{L}^\#$ of the Laplacian of $\hat{\cal G}$ and the inverse $C$ of the matrix $\hat L+ {\bf 1}_n{\bf 1}^{\rm T}_n$, whose entries are used to compute the information $I_{ij}$ via \eqref{eq:combined_path_info_matrix}. 

\begin{lemma} \label{lem:group_inverse_and_info_matrix}
Let $\hat{\cal G}$ be an undirected connected graph of order $n$ with Laplacian matrix $\hat L$. Then,
\begin{equation} \label{eq:TAC10:information_matrix}
C = (\hat L +  {\bf 1}_n{\bf 1}^{\rm T}_n)^{-1} = \hat L^\# + \frac{1}{n^2}  {\bf 1}_n{\bf 1}^{\rm T}_n,
\end{equation}
where $\hat L^\#$ denotes the group inverse of $\hat L$.
\end{lemma}

\begin{IEEEproof}
The result follows from  
\begin{equation}\nonumber
\begin{aligned}
( \hat L + {\bf 1}_n{\bf 1}^{\rm T}_n ) \left(\hat L^\# + \frac{1}{n^2} {\bf 1}_n{\bf 1}^{\rm T}_n \right) &= \hat L \hat L^\# + \frac{1}{n^2} \left(\hat L{\bf 1}_n \right) {\bf 1}^{\rm T}_n \\
&+ {\bf 1}_n \left( {\bf 1}^{\rm T}_n \hat L^\# \right) + \frac{1}{n}{\bf 1}_n{\bf 1}^{\rm T}_n\\
%
& = I_n,
\end{aligned}
\end{equation}
where Lemma \ref{lem:laplacian_group_inverse_properties} and the fact ${\bf 1}^{\rm T}_n{\bf 1}_n=n$ have been used.
\end{IEEEproof}

Theorem \ref{thm:TAC10:information_centrality_ranking} is now proved by combining the lemmas above. 
\begin{IEEEproof}
The definition of information centrality \eqref{eq:info_centrality} combined with \eqref{eq:combined_path_info_matrix} gives
\begin{equation}
\frac{1}{\hat \kappa_{\rm {info}}(v_k)} =  \frac{1}{n} \sum^n_{j=1} (c_{kk} + c_{jj} - 2c_{kj}).
\end{equation}
By \eqref{eq:TAC10:information_matrix} in view of \eqref{eq:g_inverse_diagonal} we have
\begin{equation}
\frac{1}{n} \sum^n_{j=1} c_{kk} = c_{kk} = \hat L^\#_{kk} + \frac{1}{n^2} = \frac{2}{\sigma^2} \frac{1}{\mu(v_k)} + \frac{1}{n^2}.
\end{equation}
In addition, by \eqref{eq:TAC10:information_matrix} and the definition of Kirchhoff index \eqref{eq:kirchhoff_index}, 
\begin{equation}
\frac{1}{n} \sum^n_{j=1} c_{jj} = \frac{1}{n}{\rm {Tr}}(\hat L^\#) + \frac{1}{n^2} = \frac{1}{n^2} \hat K_{\rm f} + \frac{1}{n^2}
\end{equation}
and
\begin{equation}
\frac{1}{n} \sum^n_{j=1} 2c_{kj} = \frac{2}{n} \left( \sum^n_{j=1} \hat L^\#_{kj} + \frac{1}{n} \right) = \frac{2}{n^2},
\end{equation}
where \eqref{eq:TAC10:laplacian_group_inverse_properties_2} was used. The result follows. 
\end{IEEEproof}

\section{Classes of Normal Graphs}
\label{sec:TAC10:examples_normal}

This section examines how the node classification based on the certainty index $\mu$ depends on certain graph parameters, e.g. the order of the graph, for some common families of graphs.

\begin{figure}[b!]
\vspace{-0.35in}
\centerline{
    \includegraphics[width=0.65\columnwidth]{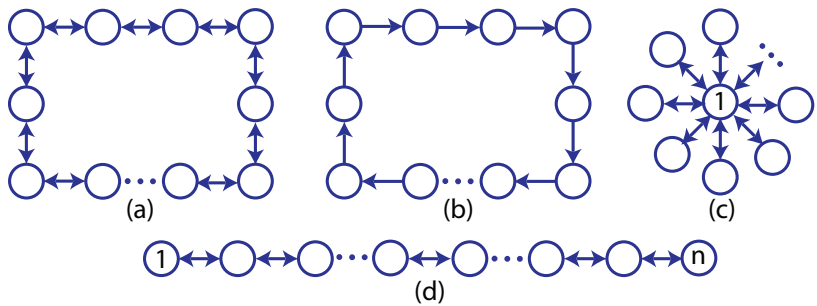}}
\caption{Examples of normal graphs. (a) An undirected ring. (b) A directed ring. (c) An undirected star. (d) An undirected path. The undirected and the directed rings are examples of circulant graphs.}
\label{fig:normal_graphs}
\end{figure}

\subsection{$d_0$-Circulant Graphs}
\label{subsubsec:TAC11:circulant_graph}

In these graphs, each node is connected to $d_0$ other nodes, where $d_0$ is a fixed integer in the interval $[2, n-1]$. For example, the complete graph is $(n-1)$-circulant and the undirected ring is $2$-circulant; see Fig. \ref{fig:normal_graphs}(a). The corresponding Laplacian can be completely determined by its first row; In more detail, if $\{l_0, l_1, ... l_{n-1}\}$ are the elements of the first row of $L$, then
\begin{equation}
L_{k,j} = l_{(k-j)~{\rm {mod}}~n}.
\end{equation}
Then, in view of results in \cite{DA79}, Lemma \ref{lem:normal_var} gives the following expression for the node certainty index
\begin{equation}\label{eq:circ_mu}
\frac{1}{\mu(v_k)} = \sigma^2 \frac{1}{n} \sum^{n}_{p=2} \frac{1}{2 {\rm {Re}}(\lambda_p)},
\end{equation}
where $\lambda_p$ denotes the $p$-th eigenvalue of $L$,
\begin{equation} \label{eq:circ_eigenvalues}
    \displaystyle \lambda_p = \sum^{n-1}_{q=0} l_q e^{-i \frac{2 \pi q (n+1-p)}{n}}.
\end{equation}
Clearly, \eqref{eq:circ_mu} indicates that the value $\mu(v_k)$ does not depend on $v_k$, implying that
\begin{equation}
\mu(v_1) = \mu(v_2) = ... = \mu(v_n).
\end{equation}
Hence, in such graphs, no distinction can be made among the nodes in terms of the variance of their state.

In the particular case of a complete graph on $n$ vertices, and assuming that $\sigma=1$ and that the weight of all the edges is equal to $\alpha$, the corresponding Laplacian $L$ is 
\begin{equation} \label{eq:TAC11:laplacian_complete}
    L= \alpha \left( n I_n - {\bf 1}_n {\bf 1}_n^{\rm T} \right),
\end{equation}
from which a computation detailed in Appendix \ref{app:TAC:complete} results in
\begin{equation}\label{eq:TAC11:mu_complete}
\mu(v_k) = 2\alpha \frac{n^2}{n-1}.
\end{equation}
Thus, for a given coupling strength $\alpha$, the index $\mu(v_k)$ increases linearly with the order $n$ for the complete graph. This implies that, after a transient, the variance of \emph{all} nodes can be made arbitrarily close to the minimum achievable variance $t/n$ by increasing the order of the graph\footnote{This observation, however, does not extend to all graphs with circulant Laplacians. A counterexample is provided by the unweighted undirected rings, in which the variance is bounded away from zero as the order $n$ increases.}. 

\subsection{Undirected Star}
\label{subsubsec:TAC11:undirected_star}

Assuming that $v_1$ is the center of the star (the highest degree node) as in Fig. \ref{fig:normal_graphs}(c), and that the weight of all edges is equal to $\alpha$ and $\sigma=1$, the corresponding Laplacian is
\begin{equation} \label{eq:laplacian_star_bi}
    L=
    \begin{bmatrix}
    (n-1)\alpha & -\alpha ~{\bf 1}^{\rm T}_{n-1} \\
    -\alpha {\bf 1}_{n-1} & \alpha I_{n-1}
    \end{bmatrix}.
\end{equation}
The structure of the Laplacian allows for an explicit computation of the covariance matrix; see Appendix \ref{app:TAC:star_bi} for details. 

By \eqref{eq:TAC10:mu_definition} the resulting expressions for $\mu$ are 
{\setlength\arraycolsep{0.1em}
\begin{eqnarray} \label{eq:TAC10:mu_undir_star}
\mu(v_1) &=& \alpha~\frac{2n^2}{n-1}, \\
\mu(v_k) &=& \alpha~\frac{2 (n-1) n^2}{n^3-2 n^2 +1},
\end{eqnarray}}
\noindent for $k=2,...,n$. Clearly, since
\begin{equation}
\mu(v_k) = \frac{n-1}{n^2-n-1} \mu(v_1)
\end{equation}
for all $k=2,...,n$, we have $\mu(v_1) \geq \mu(v_k)$ for undirected stars of order $n \geq 2$. 
%

Due to the explicit form of $\mu$, a number of interesting observations can be made. First, for a fixed $\alpha$, by increasing the number of nodes $n$ the index $\mu(v_1)$ increases without bound. This reflects the fact that ${\rm {Var}}(x_1)$ can be made arbitrarily close to the minimum possible variance $t / n$. On the other hand, increasing $n$ also increases $\mu(v_k)$, which though is upper bounded by $2 \alpha$. Hence, contrary to the center $v_1$, the variance ${\rm {Var}}(v_k)$ for $k=2,...,n$ cannot be made arbitrarily close to $t / n$ by increasing $n$. 
However, it is easy to see that 
\begin{equation}
\begin{aligned}
 \frac{1}{\mu(v_k)} - \frac{1}{\mu(v_1)} = \left(1-\frac{2}{n}\right) \frac{1}{2\alpha}, 
\end{aligned}
\end{equation}
which implies that as the coupling strength $\alpha$ increases the center essentially becomes equivalent to the rest of the nodes. In other words, a high value of $\alpha$ brings the nodes ``closer''. 
%


\subsection{Undirected Path}
\label{subsubsec:TAC11:undirected_path}

Consider an undirected path of order $n \geq 2$ as in Fig. \ref{fig:normal_graphs}(d). We assume that $\sigma=1$ and that all edges carry the same weight $\alpha$. The corresponding Laplacian is
\begin{equation} \label{eq:laplacian_path_undir}
    L=
    \begin{bmatrix}
    \alpha  & -\alpha & 0 & 0 & \cdots & 0 & 0 & 0\\
    -\alpha & 2\alpha & -\alpha & 0 & \cdots & 0 & 0 & 0\\
    0 & -\alpha & 2\alpha & -\alpha & \cdots & 0 & 0 & 0 \\
    \vdots & \vdots & \vdots & \vdots & \ddots & \vdots & \vdots \\
    0 & 0 & 0 & 0 & \cdots & -\alpha & 2\alpha & -\alpha \\
    0 & 0 & 0 & 0 & \cdots & 0 & -\alpha & \alpha
    \end{bmatrix},
\end{equation}
which has the structure of a symmetric tridiagonal matrix. Closed-form expressions for the eigenstructure of such matrices can be found; see Lemma \ref{lem:tridiag_eigen} in Appendix \ref{app:path_undir}. As a result, 
using \eqref{eq:TAC10:mu_definition} we can find 
\begin{equation}\label{eq:mu_path_undir}
\frac{1}{\mu(v_k)} = \frac{1}{2\alpha n} \sum^n_{p=2} \frac{\cos^2\left[\frac{\pi}{n} (p-1)(k-\frac{1}{2}) \right]}{\left(1 - \cos\left[\frac{\pi}{n}(p-1)\right] \right)}.
\end{equation}

From \eqref{eq:mu_path_undir} it can be seen that nodes symmetrically located with respect to the midpoint of the path, i.e., the pairs $(k, n-k+1)$ for $k = 1, 2, ..., n$, exhibit the same certainty in collecting evidence; see Appendix \ref{app:path_undir}. Moreover, the closer a node is to the midpoint of the path, the higher is its certainty index; equivalently, the closer is its variance to $t/n$.

Finally, it is evident from \eqref{eq:mu_path_undir} that increasing the number of nodes $n$ results in higher values of $\mu$, so that the variance of each node can be made arbitrarily close to the minimum possible variance $t/n$. In addition, the limiting case of strong communication, i.e. $\alpha \to \infty$, has the same effect as that in the undirected star, i.e., it brings node certainties closer.
 

\subsection{Comparison Across Classes of Normal Graphs}
\label{subsubsec:comparison}
\begin{figure}[b!]
\vspace{-0.5in}
\centerline{
    \subfigure[]{
    \includegraphics[width=0.49\columnwidth]
    {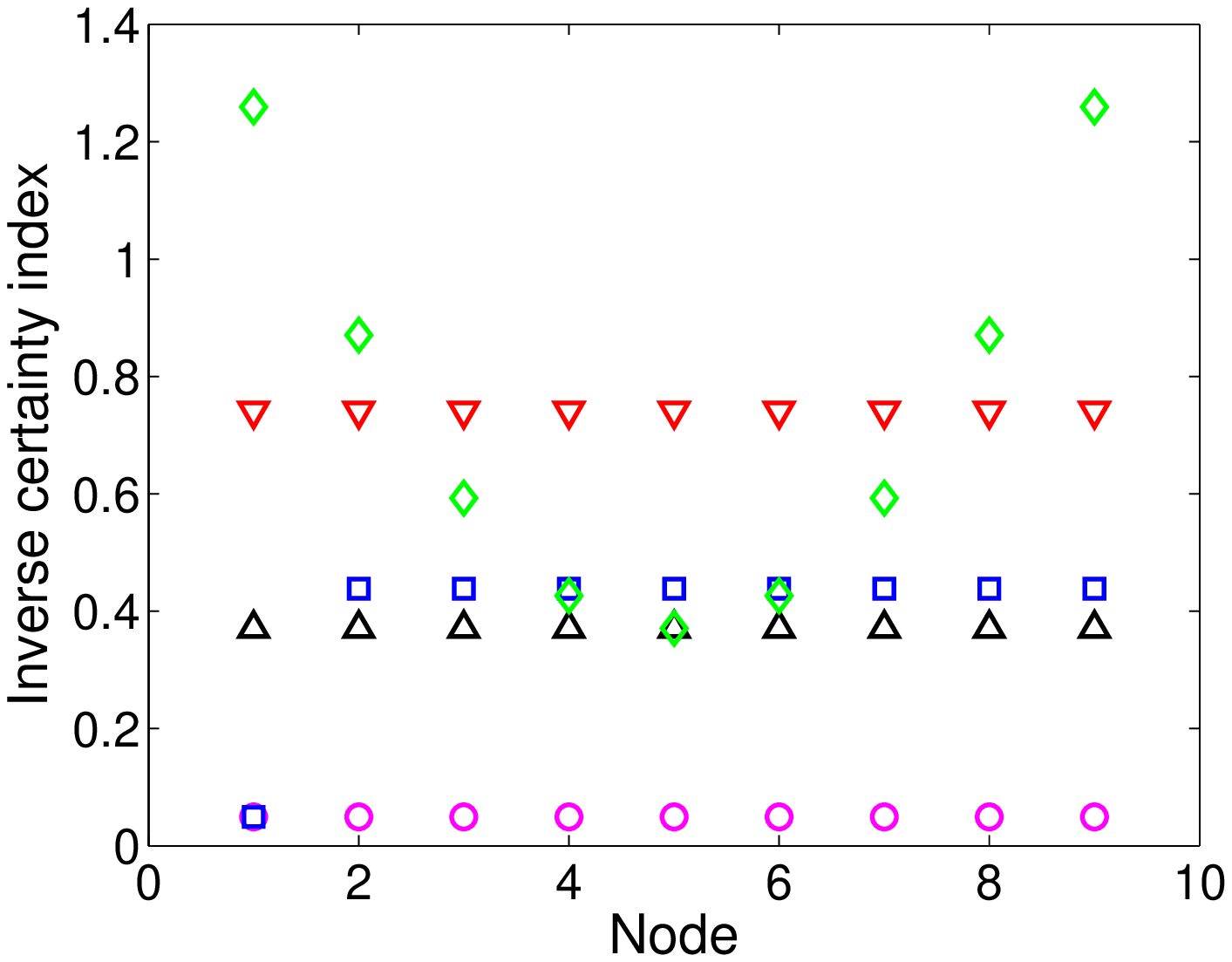}
    \label{fig:compare_mu}}
    \hfil
    \subfigure[]{
    \includegraphics[width=0.49\columnwidth] 
    {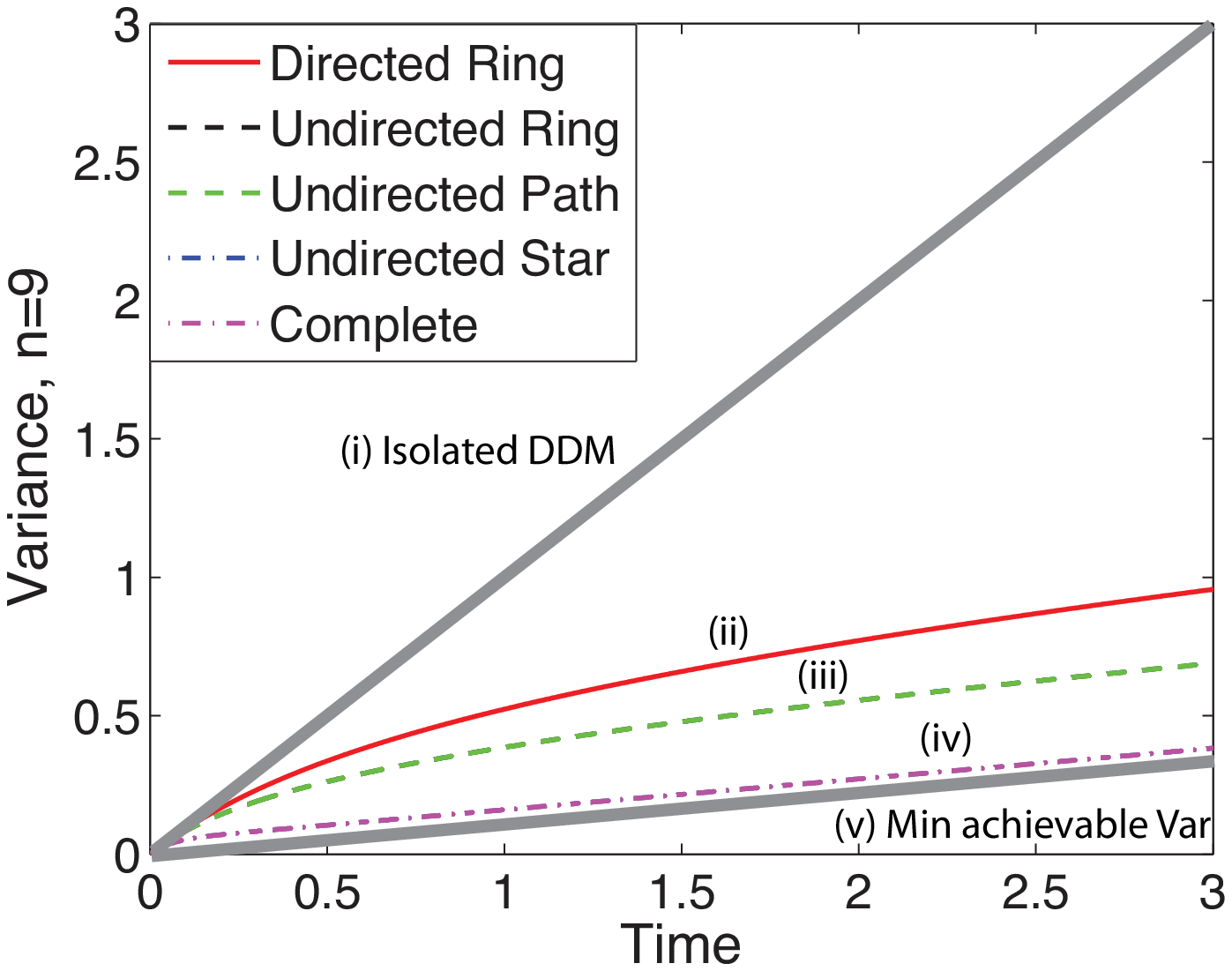}
    \label{fig:compare_var}}
    }
\caption{(a) Inverse certainty index for each node in the directed ring ($\bigtriangledown$, red), undirected ring ($\bigtriangleup$, black), undirected star ($\square$, blue), undirected path ($\diamond$, green), and complete graph ($\circ$, magenta). The best performance corresponds to the complete graph, and it is matched by the center of the undirected star. (b) Minimum variance for the graphs of Fig. \ref{fig:compare_mu}. The grey lines (i) and (v) correspond to the variance of an isolated DDM ($\sigma^2 t$) and the minimum achievable variance ($\sigma^2 t / n$), respectively; all graphs are between these lines. The variance of the nodes in the undirected ring coincide with the minimum variance node of the undirected path, both represented by curve (iii); see also Fig. \ref{fig:compare_mu}. Finally, the variance of the nodes in the complete graph coincides with the variance of the center of the undirected star, both corresponding to curve (iv); see also Fig. \ref{fig:compare_mu}. The nodes in the complete graph and the center of the undirected star are the closest to the minimum possible variance $\sigma^2 t / n$.}
\label{fig:compare_graph}
\end{figure}

Fig. \ref{fig:compare_mu} presents the inverse of the certainty index $1/\mu$ for three examples of circulant graphs---namely, the complete graph and the undirected and directed rings---and the undirected star and path graphs, all of order $n=9$; see also Fig. \ref{fig:normal_graphs}.
%
It is evident that the nodes of graphs with circulant Laplacian matrices are all equivalent in terms of their certainty. Furthermore, the nodes of the complete graph have the least uncertainty, since the variances of their states are closer to the minimum possible variance. Notice also that the center of an undirected star exactly matches the performance of the nodes of a complete graph; this can be verified by inspecting \eqref{eq:TAC11:mu_complete} and \eqref{eq:TAC10:mu_undir_star}. Similarly, the center of the undirected path matches the performance of the nodes in an undirected ring.

Fig. \ref{fig:compare_var} presents the evolution in time of the variance associated with the state of the node that maximizes $\mu$ in each of the graphs discussed above. Consistent with the results of Fig. \ref{fig:compare_mu}, the variance of the center of the undirected star coincides with that of the nodes in the complete graph, and they both evolve closely to the minimum achievable variance $\sigma^2 t /n$; see curve (iv) in Fig.\ref{fig:compare_var}. The same holds for the center of the undirected path and the nodes of the undirected ring; see curve (iii) in Fig.\ref{fig:compare_var}. Fig. \ref{fig:compare_var} illustrates that the index $\mu$ effectively ``compresses'' the evolution of the variance in time to a scalar that can be used to rank the nodes in terms of their certainty and compare the graphs accordingly.


\section{Examples of Non-normal Graphs}
\label{sec:TAC10:non_normal}

In this section, two examples are presented of digraphs with Laplacians that do not meet the normality condition of Theorem \ref{thm:TAC10:information_centrality_ranking}, but for which we can still compute the variance, an thus the certainty, of each node. These are the exploding and imploding stars, which are graphs commonly used in decentralized decision making \cite{TS93}, and consensus protocols \cite{Wu08}, respectively.

\begin{figure}[b!]
    \center
    \vspace{-0.4in}
    \includegraphics[width=0.6\columnwidth]{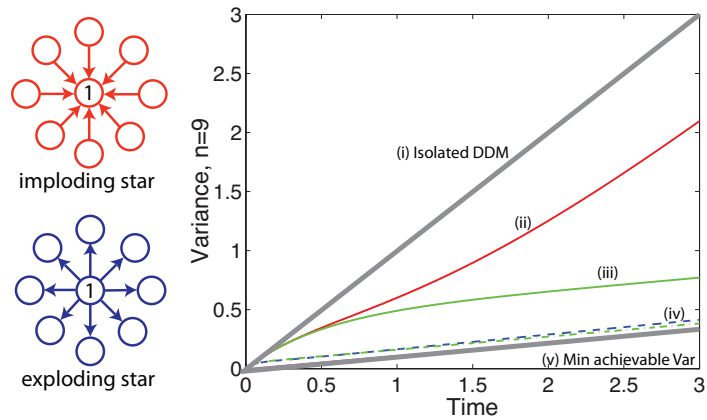}      
    \vspace{-0.2in}
    \caption{Left: Imploding and exploding stars for $n=9$. Right: Time evolution of the variances of the nodes of the stars on the left compared with undirected stars of the same order. From top to bottom: (i) variance of an isolated DDM, $\sigma^2 t$, which coincides with ${\rm {Var}}(x_k(t))$ for $k=2,...,n$ for the exploding star and with ${\rm {Var}}(x_1(t))$ for the imploding star; (ii) ${\rm {Var}}(x_k(t))$ for $k=2,...,n$ for the imploding star; (iii) ${\rm {Var}}(x_k(t))$ for $k=2,...,n$ for the undirected star (plotted for comparison); (iv) the variances of the centers of the undirected and exploding stars almost coincide; and (v) minimum achievable variance $\sigma^2 t / n$.}
    \label{fig:compare_var_star}
\end{figure}

\begin{example}[exploding star]
\label{ex:TAC11:star_expl}
The exploding star of order $n$ is considered; Fig. \ref{fig:compare_var_star}(left) presents an exploding star for $n=9$. Its Laplacian, assuming equal weight $\alpha$ assigned to all edges, is 
\begin{equation} \label{eq:laplacian_star_expl}
    L=
    \begin{bmatrix}
    (n-1)\alpha & -\alpha ~{\bf 1}^{\rm T}_{n-1} \\
    0 & 0
    \end{bmatrix}.
\end{equation}
A computation presented in Appendix \ref{app:star_expl} reveals that the variance of the center is 
\begin{equation}\label{eq:var_star_expl_center}
\begin{aligned}
    {\rm {Var}}(x_1(t)) &= -\frac{2}{(n-1)^2\alpha}+\frac{t}{n-1}+\frac{2}{(n-1)^2\alpha} e^{-(n-1) \alpha t} \\
    &+ \frac{n}{2 (n-1)^2 \alpha} \Big(1-e^{-2(n-1) \alpha t} \Big),
\end{aligned}
\end{equation}
while the variance of all the other nodes equals that of an isolated DDM, i.e., ${\rm {Var}}(x_k(t))=\sigma^2 t, ~k=2,...,n$. 

As expected, ${\rm {Var}}(x_k(t)) > {\rm {Var}}(x_1(t))$ for all $k \neq 1$ and $t > 0$, since the center has access to the information collected by the rest of the nodes, which operate in isolation from each other and from the center. Note that this relative hierarchy of ``informed/uninformed'' nodes---with the informed node being the center---persists\footnote{This should be contrasted with the behavior of the nodes of undirected stars, in which as $\alpha$ increases all nodes become equivalent; see Section \ref{subsubsec:TAC11:undirected_star}.} as the coupling strength $\alpha$ increases, and in the limit as $\alpha$ tends to infinity, the state $x_1$ of the center converges with probability one to the \emph{average} of the states of the rest of the nodes; see Appendix \ref{app:star_expl}. 
As a final remark note that, for finite $t$ and $\alpha$, the larger is the number $n$ of nodes, the smaller is ${\rm {Var}}(x_1(t))$; this implies that the uncertainty associated with the state of the center can be made arbitrarily small by increasing $n$. 


\end{example}


\begin{example}[imploding star]
\label{ex:TAC11:star_impl}
Consider an imploding star of order $n$; see Fig. \ref{fig:compare_var_star} (left) for the imploding star of order $n=9$. Its Laplacian, assuming equal weight $\alpha$ assigned to all edges, is
\begin{equation} \label{eq:laplacian_star_impl}
    L=
    \begin{bmatrix}
    0 & 0 \\
    -\alpha {\bf 1}_{n-1} & \alpha I_{n-1}
    \end{bmatrix}.
\end{equation}
The variance of the center is ${\rm {Var}} (x_1(t)) = \sigma^2 t$, equal to that of an isolated drift-diffusion process, and 
\begin{equation} \label{eq:cov_star_impl}
{\rm {Var}} (x_k(t)) =  t + \frac{2 e^{-\alpha t}-e^{-2 \alpha t} -1}{\alpha},
\end{equation}
for $k=2,...,n$; see Appendix \ref{app:star_impl}.

Thus, the variance associated with the state of the center is always larger than the variance associated with the state of each of the other nodes. The difference between ${\rm {Var}}(x_1(t))$ and ${\rm {Var}}(x_k(t)) $ for $k \in \{2,...,n\}$ as $t$ grows eventually approaches the constant $1 / \alpha$, which decreases as the coupling strength $\alpha$ increases. Hence, for strong coupling, the variance associated with each node deteriorates, approaching $\sigma^2 t$. 
Most important---contrary to all the graphs studied so far---the variances of the nodes of an imploding star are all \emph{independent} of its order $n$. This is because information flows to the nodes from the center but not in the other direction. Hence, the pairs formed by each node together with the center are decoupled.
\end{example}

Example \ref{ex:TAC11:star_expl} shows that the center of the exploding star nearly achieves the performance of the nodes of a complete graph in terms of certainty; see Fig. \ref{fig:compare_var} and  Fig. \ref{fig:compare_var_star}. It is known however that exploding stars cannot achieve consensus because they are not connected. In the reverse case, the imploding star provides fast rate of convergence to consensus \cite[Thm. 2.38]{Wu08} in a deterministic setting and strong robustness of consensus to additive (white) noise in a stochastic setting \cite[Fig. 1]{YoScLe10}, approaching the behavior of a complete graph as its order increases. However, as Example \ref{ex:TAC11:star_impl} and Fig. \ref{fig:compare_var_star} reveal, the imploding star exhibits the worst performance in terms of the certainty of its nodes and the variance associated with their states cannot be affected by increasing the order $n$ of the graph. These examples reveal a fundamental difference in the mechanisms that underlie robustness of consensus to noise and individual unit certainty in evidence accumulation.

\section{Conclusion}
\label{sec:TAC10:conclusions}

This paper investigated the effect of the communication graph on the certainty across a network of stochastic evidence accumulators. Each unit in the network accumulates evidence according to a drift-diffusion process based on its own noisy measurements and on evidence exchanged with other units according to the communication graph. The use of a drift-diffusion model to represent evidence accumulation was motivated by a class of collective decision-making problems. It was observed that the uncertainty in the network depends on the Laplacian of the communication graph, implying that certain nodes within a given communication topology can be more certain than others. To classify the nodes according to their certainty in accumulating evidence, a node certainty index was defined based on the eigenstructure of the graph's Laplacian. It was proved that the certainty index can be interpreted as a graph centrality measure through the notion of information centrality, which incorporates \emph{all} possible paths connecting each unit with the rest of the network. These results show that evidence accumulation is a total network process, and can be used to identify units that are more influential than others in tasks that involve translating the accumulated evidence to decisions.


%

\appendix

\subsubsection{All-to-all Communication}
\label{app:TAC:complete}

The $k$-th power of the Laplacian defined in \eqref{eq:TAC11:laplacian_complete} is by induction to be
\begin{equation}\label{eq:laplacian_powers_all}
    L^k(t-\tau)^k = \alpha^k(t-\tau)^k \left(n^k I - n^{k-1} {\bf 1}{\bf 1}^{\rm T} \right).
\end{equation}
Substitution of \eqref{eq:laplacian_powers_all} to the matrix exponential series gives
\begin{equation}\label{eq:laplacian_exp_all}
    e^{-L(t-\tau)} = e^{-n \alpha (t -\tau)} I + \frac{1 - e^{n \alpha (t - \tau)}}{n} {\bf 1} {\bf 1}^T.
\end{equation}
from which, through \eqref{eq:SDE_solution_cov}, we obtain


\begin{equation} \nonumber
    {\rm {Cov}} (x(t), x(t)) = \sigma^2 \left[ \frac{1-e^{-2 n \alpha t}}{2 n \alpha}~I
    +\left( \frac{t}{n} - \frac{1 - e^{-2 n \alpha t}}{2 n^2 \alpha} \right) {\bf 1} {\bf 1}^{\rm T} \right].
\end{equation}

\subsubsection{Undirected Star Topology}
\label{app:TAC:star_bi}

By induction, the $k$-th power of the Laplacian of an undirected star of order $n$ is
\begin{equation}\label{eq:laplacian_powers_star_undir}
    L^k = \alpha^k
    \begin{bmatrix}
        n^{k-1}(n-1) & -n^{k-1} {\bf 1}_{n-1}^{\rm T} \\
        -n^{k-1} {\bf 1} & I + \left( \sum^{k-2}_{j=0}n^j \right) {\bf 1}{\bf 1}^{\rm T}
    \end{bmatrix},
\end{equation}
where ${\bf 1}$ corresponds to an $(n-1)$-dimensional column vector of ones and $I$ is the $(n-1) \times (n-1)$ identity matrix. Substitution of \eqref{eq:laplacian_powers_star_undir} in the matrix exponential series gives
\begin{equation}\label{eq:laplacian_exp_star_bi}
    e^{-L(t-\tau)} =
    \begin{bmatrix}
        e_1(t, n, \alpha) & e_2(t, n, \alpha) {\bf 1}^{\rm T} \\
        e_2(t, n, \alpha) {\bf 1}_{n-1} & e_3(t, n, \alpha) I + e_4(t, n, \alpha) {\bf 1}{\bf 1}^{\rm T}
    \end{bmatrix},
\end{equation}
where
\begin{equation}
\begin{array}{ll}
    e_1(t, n, \alpha) = \frac{1}{n} + \frac{n-1}{n} e^{-n \alpha (t-\tau)}, &
    e_2(t, n, \alpha) = \frac{1}{n} - \frac{e^{-n \alpha (t-\tau)}}{n}, \\
    e_3(t, \alpha) = e^{- \alpha (t - \tau)}, &
    e_4(t, n, \alpha) = \frac{1}{n} - \frac{e^{-\alpha (t - \tau)}}{n-1} + \frac{e^{- n \alpha (t-\tau)}}{n(n-1)}.
\end{array}
\end{equation}
Substituting these expressions in \eqref{eq:SDE_solution_cov} we obtain
\begin{equation} \label{eq:cov_star_bi}
{\rm {Cov}} (x(t), x(t)) = \sigma^2
\begin{bmatrix}
    c_1 & c_2~{\bf 1}^{\rm T} \\
    c_2~{\bf 1} & c_3~I + c_4 {\bf 1}{\bf 1}^{\rm T}
\end{bmatrix},
\end{equation}
in which
\begin{equation}\label{eq:cov_star_bi_entries}
\begin{array}{ll}
    c_1(t, n,\alpha) = \frac{t}{n} + \frac{n-1}{2 \alpha n^2} \left( 1 - e^{-2 n \alpha t} \right), &
    c_2(t, n,\alpha) = \frac{t}{n} - \frac{1-e^{-2 n \alpha t}}{2 n^2 \alpha}, \\
    c_3(t, \alpha) = \frac{1-e^{- 2 \alpha t}}{2 \alpha}, &
    c_4(t, n,\alpha) = \frac{t}{n} - \frac{1-e^{-2 \alpha t}}{2 (n-1) \alpha} + \frac{1-e^{- 2 n \alpha t}}{2 n^2 (n-1) \alpha}.
\end{array}
\end{equation}

\subsubsection{Undirected Path Topology}
\label{app:path_undir}

The Laplacian \eqref{eq:laplacian_path_undir} is a symmetric tridiagonal matrix and the following lemma applies.
\begin{lemma}[\cite{CHetal09}] \label{lem:tridiag_eigen}
Let $L$ be the Laplacian matrix \eqref{eq:laplacian_path_undir}. Then, for $p = 1, 2, ..., n$,
\begin{equation} \label{eq:tridiag_evalue}
    \lambda_p = 2 \alpha \left( 1 - \cos \left[ \frac{\pi}{n} (p-1) \right]\right)
\end{equation}
is an eigenvalue of $L$, and
\begin{eqnarray} \label{eq:tridiag_evector}
  u^{(p)}_k =
  \begin{cases}
    \begin{aligned}
        \frac{1}{\sqrt{n}}, & ~~p=1\\
        \frac{2}{\sqrt{n}} \cos \left[ \frac{\pi}{n}(p-1)(k-\frac{1}{2})\right], & ~~p = 2, .., n.
    \end{aligned}
  \end{cases}
\end{eqnarray}
is the $k$-th, $k \in \{1, \ldots, n\}$ , component of the corresponding normalized $p$-th eigenvector. 
\end{lemma}

In view of Lemma \ref{lem:tridiag_eigen}, Lemma \ref{lem:normal_var} gives
\begin{equation}
\begin{aligned}
    \left[{\rm {Cov}}(x(t), x(t))\right]_{k j} &= \sigma^2 \left( \frac{t}{n} + \frac{2}{n} \sum^{n}_{p=2} \frac{1 - e^{-\lambda_p t}}{2 \lambda_p}  \zeta^{(p)}_{kj} \right)
\end{aligned}
\end{equation}
where
\begin{equation}
 \zeta^{(p)}_{k,j}=\cos \left[ \frac{\pi}{n}(p-1)(k-\frac{1}{2})\right] \cos \left[ \frac{\pi}{n}(p-1)(j-\frac{1}{2})\right].
\end{equation}
Then, the variance of each node is readily obtained for $j=k$, resulting in \eqref{eq:mu_path_undir}. Finally, to show that the variances of the nodes that are symmetrically located with respect to the midpoint of the path are equal, note that
\begin{equation}
 \zeta^{(p)}_{k, k} =  \zeta^{(p)}_{n-k+1, n-k+1}
\end{equation}
due to the fact that
\begin{equation}
\cos^2 \left[ \frac{\pi}{n}(p-1)(k-\frac{1}{2})\right] - \cos^2 \left[ \frac{\pi}{n}(p-1)(n-k+\frac{1}{2})\right] = 0.
\end{equation}

\subsubsection{Exploding Star Topology}
\label{app:star_expl}

The $k$-th power of the Laplacian of the exploding star is found by induction to be
\begin{equation}\label{eq:laplacian_powers_star_expl}
    L^k = \alpha^k
    \begin{bmatrix}
        (n-1)^k & -(n-1)^{k-1} {\bf 1}\\
        0 & 0
    \end{bmatrix}.
\end{equation}
Substitution of \eqref{eq:laplacian_powers_star_expl} in the matrix exponential series results in
\begin{equation}\label{eq:laplacian_exp_star_expl}
    e^{-L(t-\tau)} =
    \begin{bmatrix}
        e^{-(n-1) \alpha (t-\tau)} & \frac{1}{n-1} \left( 1-e^{-(n-1) \alpha (t-\tau)} \right) {\bf 1}\\
        0 & I
    \end{bmatrix}.
\end{equation}
Substitution of \eqref{eq:laplacian_exp_star_expl} in \eqref{eq:SDE_solution_cov} results in,
\begin{equation} \label{eq:cov_star_expl}
{\rm {Cov}} (x(t), x(t)) = \sigma^2
\begin{bmatrix}
    c_1 & c_2~{\bf 1}^{\rm T} \\
    c_2~{\bf 1} & t~I
\end{bmatrix},
\end{equation}
where
\begin{equation}\label{eq:cov_star_expl_entries}
\begin{aligned}
    c_1(t, n,\alpha) &= -\frac{2}{(n-1)^2\alpha}+\frac{t}{n-1}+\frac{2}{(n-1)^2\alpha} e^{-(n-1) \alpha t} \\
    &+ \frac{n}{2 (n-1)^2 \alpha} \Big(1-e^{-2(n-1) \alpha t} \Big), \\
    c_2(t, n,\alpha) &= -\frac{1}{(n-1)^2 \alpha} + \frac{t}{n-1} + \frac{1}{(n-1)^2 \alpha} e^{-(n-1) \alpha t}.
\end{aligned}
\end{equation}

Note that for fixed $t$ and $n$ \eqref{eq:cov_star_expl_entries} implies that, 
\begin{equation} \label{eq:cov_star_expl_lim_alpha}
\lim_{\alpha \to \infty}{\rm {Cov}} (x(t), x(t)) ~= \sigma^2
\begin{bmatrix}
    \frac{t}{n-1} & \frac{t}{n-1}~{\bf 1}^{\rm T} \\
    \frac{t}{n-1}~{\bf 1} & t~I
\end{bmatrix},
\end{equation}
and the limiting covariance matrix is singular with ${\rm {dim}}[{\cal N} \left( {\bf K}(t) \right)] = 1$. To provide insight into the singular nature of the random vector $x(t)$, consider the new variable
\begin{equation} \label{eq:star_expl_new_var_alpha_lim}
    y = x_1 - \frac{1}{n-1} \sum^n_{j=2} x_j,
\end{equation}
where $x_1$ is the state of the informed node. Then, $\mathbb{E}[y(t)] = 0$ and ${\rm {Var}}(y(t)) = 0$ meaning that $y$ is deterministic.

\subsubsection{Imploding Star Topology}
\label{app:star_impl}

Observe that $L^2 = \alpha^2 L$, i.e., the Laplacian of the imploding star is an indempotent matrix. Hence, the $k$-th power of $L$ is
\begin{equation}\label{eq:laplacian_powers_star_impl}
    L^k = \alpha^k
    \begin{bmatrix}
        0 & 0 \\
        -{\bf 1} & I
    \end{bmatrix},
\end{equation}
from which
\begin{equation}\label{eq:laplacian_exp_star_impl}
    e^{-L(t-\tau)} =
    \begin{bmatrix}
        1 & 0\\
       \left( 1-e^{-\alpha (t-\tau)} \right){\bf 1} & e^{-\alpha(t-\tau)} I
    \end{bmatrix}.
\end{equation}
Substitution in \eqref{eq:SDE_solution_cov} leads to
\begin{equation} \label{eq:cov_star_impl_app}
{\rm {Cov}} (x(t), x(t)) = \sigma^2
\begin{bmatrix}
    t & c_1 {\bf 1}^{\rm T} \\
    c_1 {\bf 1} & c_2 {\bf 1}{\bf 1}^{\rm T} + c_3 I
\end{bmatrix},
\end{equation}
where
\begin{equation}\label{eq:cov_star_impl_entries}
    c_1(t, \alpha) = t-\frac{1-e^{-\alpha t}}{\alpha}, ~
    c_2(t, \alpha) = -\frac{3}{2 \alpha} + t + \frac{2 e^{-\alpha t}}{\alpha} - \frac{e^{-2 \alpha t}}{2 \alpha},~
    c_3(t, \alpha) = \frac{1}{2 \alpha} - \frac{e^{- 2 \alpha t}}{2 \alpha}.
\end{equation}

\bibliographystyle{IEEEtran}
\bibliography{decision_graph}

\end{document}